\documentclass[trackchanges]{aastex702}

\renewcommand{\added}[1]{\textcolor{blue}{\textbf{#1}}}
\renewcommand{\added}[1]{#1}
\usepackage{amsmath}
\usepackage{multirow}
\usepackage{rotating}
\usepackage{url}
\usepackage{hyperref}

\usepackage{listings,xcolor}
\usepackage{listings,xcolor}
\definecolor{codebg}{gray}{0.95}      
\definecolor{codeframe}{gray}{0.80}   

\begin{document}

\title{
\texttt{CARMApy}: An Open-Source Python Framework for Simulating Microphysical Clouds in Planetary Atmospheres}

\author[0000-0002-8658-3811]{Wolf Cukier}
\affiliation{Department of Astronomy \& Astrophysics, the University of Chicago, Chicago, IL, 60637, USA}
\email[show]{wcukier@uchicago.edu}  

\author[orcid=0000-0002-4250-0957]{Diana Powell} 
\affiliation{Department of Astronomy \& Astrophysics, the University of Chicago, Chicago, IL, 60637, USA}
\email{}

\author[0000-0002-8706-6963]{Xi Zhang}
\affiliation{Department of Earth and Planetary Sciences, University of California Santa Cruz, Santa Cruz, CA, 95064, USA}
\email{}

\author[0000-0002-8518-9601]{Peter Gao}
\affiliation{Earth and Planets Laboratory, Carnegie Institution for Science, Washington, DC, USA}
\email{}

\author[0000-0002-8956-2047]{Dominic Samra}
\affiliation{Department of Astronomy \& Astrophysics, the University of Chicago, Chicago, IL, 60637, USA}
\email{}

\author[0000-0001-5909-4433]{Vighnesh Nagpal}
\altaffiliation{NSF Graduate Research Fellow}
\affiliation{Department of Astronomy \& Astrophysics, the University of Chicago, Chicago, IL, 60637, USA}
\email{}

\begin{abstract}
\texttt{CARMApy} is a new open-source python code that performs bin-scheme microphysical modeling of clouds in exoplanet atmospheres.   It models key cloud properties such as particle size distributions and microphysical rates from first principles.  The code is a wrapper of \texttt{ExoCARMA}, a well tested \texttt{Fortran} code with an almost half century long heritage.  \texttt{CARMApy} includes the microphysical processes of homogeneous and heterogeneous nucleation, condensational growth, evaporation, coagulation, and vertical transport.  \texttt{CARMApy} has 11 built-in default condensates and allows the user to specify additional condensates.  In this work we describe \texttt{CARMApy} and the data products that it can generate, along with the history of its code heritage.  We additionally compile a complete description of the theory and methods used in \texttt{CARMA}.  Lastly we benchmark \texttt{CARMApy} and show that its results are consistent with previous versions of \texttt{CARMA}, while executing the code $\sim$1.9 times faster single threaded and $\sim$3.8 times faster multithreaded.

\end{abstract}



\section{Introduction} 

Clouds impart first order impacts on planetary atmospheric properties and our interpretations thereof.  Uncertainties in cloud formation processes can propagate to substantial uncertainties in inferences of atmospheric composition (eg. C/O \citep{Helling2016MineralCloudsHD}, Mg/Si \citep{Calamari2024PredictingCloudConditions}), atmospheric and interior P-T structure \citep[e.g.,][]{Molliere2020RetrievingScatteringClouds}, and atmospheric dynamics.  For example, cloud formation is sensitive to atmospheric structure and mixing in the deep atmosphere which is otherwise unobservable \citep{Powell2018FormationSilicateTitanium} and the locations of clouds in atmospheres is one of our best probes of the atmospheric dynamics on these bodies \citep[eg.][]{DiGirolamo2025DecadalChangesAtmospheric,Plummer2025MappingCloudDrivenAtmospheric}.  Clouds and radiative transfer are strongly coupled \citep{Parmentier2021CloudyShapeHot} and clouds are needed to understand planetary energy budgets \citep{Malsky2024DirectComparisonUse}.  The exact way which clouds shape these observations is determined by a number of key cloud properties such as the precise 3-D distribution of clouds on these bodies, the particle size distributions of clouds in each of these locations, the composition of the cloud particles, the shapes of the cloud particles, how all of these properties vary over time and how all these properties affect the optical properties of the cloud particles.

The field has employed models of clouds across a variety of levels of complexity that capture different amounts of these key properties.  On the simple end of this spectrum exist gray cloud (GC) models.  GC models assume a cloud deck with constant transmittance across all wavelengths, making them incredibly efficient and thus widely used in retrieval studies of exoplanet atmospheres \citep[eg. ][]{Xue2024JWSTTransmissionSpectroscopy}.  GC models make an extreme tradeoff prioritizing efficiency over capturing the physical complexities of clouds. Equilibrium cloud condensation (ECC) models \citep{Ackerman2001PrecipitatingCondensationClouds} are a class of commonly used models for exoplanetary clouds that exist a step up the complexity ladder from GC models \cite[eg.][]{Marley2021SonoraBrownDwarf, Morley2024SonoraSubstellarAtmosphere}.  ECC models, as their name implies, assume equilibrium behavior in the atmosphere---they assume that supersaturated gasses will condense out of the atmosphere, forming clouds, until the gas reaches a saturation ratio of unity. ECC modeling frameworks are fast to run, making them well suited to tasks such as 1-D models that solve for the radiative state of an atmosphere, \added{or for layering on complexity in other parts of the model such as accounting for fractal aggregates \citep{Moran2025FractalAggregateAerosols, Lodge2026FractalAggregateAerosols}}. ECC models, however, require a prescribed shape for the cloud particle size distribution and involve a fitted parameter (known as $f_\text{sed}$) with unclear physical meaning, obscuring the physical processes that might be forming these clouds. \added{State-of-the-art ECC codes such as \texttt{Virga} \citep{Batalha2026CondensationCloudsSubstellar} incorporate more physics such as calculating particle fall velocity or allowing $f_\text{sed}$ to vary with altitude \citep{Rooney2022NewSedimentationModel} but still make fundamental assumptions about the shape of the particle size distributions that might not be strictly physical.}

Moment-method (MM) models are also used to model clouds on planetary objects and fold even in more physics.  \added{These models track representative statistics of the underlying particle size distribution from first principles and calculate particle settling self-consistently.  These MM models themselves span complexity: some  such as \texttt{ARCiS} \citep{Ormel2019ARCiSFrameworkExoplanet, Min2020ARCiSFrameworkExoplanet} and \texttt{ExoLyn} \citep{Huang2024ExoLynGoldenMean} use parameterized nucleation rates for cloud formation to be efficient enough for use in retrieval frameworks; whereas others, the kinetic cloud formation models, calculate nucleation rates from first principles, exchanging speed for more physical accuracy, such as the \texttt{Drift} family of models \citep[eg. ][]{Woitke2003DustBrownDwarfs, Woitke2004DustBrownDwarfs}, \texttt{mini-cloud} \citep{Lee2023ModellingDynamicallyDriven}, and \texttt{Nimbus} \citep{Kiefer2026ConnectingJWSTSilicate}.}  \added{MM models have been used in a wide variety of contexts such as Earth and Jovian clouds \citep{Ohno2017CondensationcoalescenceCloudModel}, 1D theoretical models of brown dwarf and exoplanet atmospheres \citep{Woitke2020DustBrownDwarfs}, interpretations of JWST silicate cloud observations \citep{Kiefer2026ConnectingJWSTSilicate}, and 3D models of hot Jupiters \citep{Lee2023ModellingDynamicallyDriven}.} \added{Even though there exist improvements to these models to allow for polydisperse particle size distributions \citep{Lee2025MonoculturePolydisperseMomenta, Lee2025MonoculturePolydisperseMoment}, generating complete particle size distributions, as is required to model observables, still requires an assumed functional form.}  While all of these models have their individual strengths, inability of these models to resolve arbitrary cloud particle size distributions leads them to struggle to reproduce both certain observed spectral features and bulk properties in these bodies, especially now with the existence of high resolution JWST spectra \citep[eg. ][]{Petrus2024JWSTEarlyRelease}.  

\added{Bin-scheme microphysics models, instead of tracking the statistical properties of the underlying size distribution, track the particle size distribution itself at an arbitrary resolution.   These bin-scheme models such as the Community Aerosol and Radiation Model for Atmospheres, \texttt{CARMA} \citep{Turco1979OneDimensionalModelDescribing, Toon1988MultidimensionalModelAerosols}, and the Lavvas family of models \citep[eg.][]{Lavvas2008CouplingPhotochemistryHaze} are thus able to allow the underlying physics to dictate the emergence of arbitrary size distributions.  While this paper will focus on models descended from \texttt{CARMA}, we would like to highlight that the Lavvas family of models treats the physics in a very similar way and has been used for a number of studies such as describing evolution of high altitude photochemical hazes in hot Jupiter atmospheres \citep{Lavvas2017AerosolPropertiesAtmospheres} and modeling cloud-haze interactions on WASP-39b \citep{Arfaux2024CouplingHazeCloud}.}

\texttt{CARMA}, \added{as a bin-scheme model},  is able to naturally reproduce a number of cloud features that other \added{non bin-scheme} cloud models struggle to accurately describe such as the brown dwarf and Hot Jupiter 10 $\mu$m silicate feature \citep{Powell2019TransitSignaturesInhomogeneous}, the day/night temperature contrast in Hot Jupiter atmospheres \citep{Gao2021UniversalCloudComposition},  and the Hot Jupiter aerosol sequence \citep{Gao2020AerosolCompositionHot}. Beyond just these cases, \texttt{CARMA} is a well-tested and validated model that has been successfully applied to interpret detailed observations of clouds and aerosols both in our solar system---including sulfuric acid clouds on Venus \citep{Gao2014BimodalDistributionSulfuric}, numerous studies of Earth's clouds and aerosols \citep[eg.][]{Chen2018ImprovementStratosphericAerosol, Lian2022GlobalDistributionAsian, Rusch2017LargeIceParticles},  CO$_2$ and water clouds on Mars \citep{Michelangeli1993NumericalSimulationsFormation, Colaprete1999CloudFormationMars}, clouds and hazes on Titan \citep{Barth2003MicrophysicalModelingEthane, Barth2004PropertiesMethaneClouds, Barth2006MethaneEthaneMixed}, and hazes on Pluto \citep{Gao2017ConstraintsMicrophysicsPlutos}---and on extra-solar bodies such as Hot Jupiters \citep{Powell2018FormationSilicateTitanium, Powell2024TwodimensionalModelsMicrophysical}, sub-Neptunes \citep{Gao2023HazyMetalrichAtmosphere} and Y-dwarfs \citep{Mang2022MicrophysicsWaterClouds, Mang2024MicrophysicalPrescriptionsParameterized}.  \texttt{CARMA} is a flexible model, capable of simulating practically any condensate assuming enough is known about their physical properties.

Despite the strengths of \texttt{CARMA} as a model, it is only currently being used by a handful of researchers in the exoplanet community who have the required expertise to run the code.  As a piece of legacy scientific software written in \texttt{Fortran} and developed over almost half a century, \texttt{CARMA} requires a large time investment to understand how the software works and how to run it.  In light of this problem, we are releasing \texttt{CARMApy}---an open source python wrapper for \texttt{CARMA} which is designed to be user-friendly and well-documented in order to enable a wide range of researchers with minimal prior cloud modeling and/or \texttt{Fortran} expertise to effectively use it.   \texttt{CARMApy} is based on the \texttt{ExoCARMA} \citep[eg.][]{Powell2024TwodimensionalModelsMicrophysical, Gao2020AerosolCompositionHot} version of \texttt{CARMA 3.0} \citep{Bardeen2008NumericalSimulationsThreedimensional} and is well suited for modeling aerosol microphysics on exoplanets, as well as environments such as protoplanetary disks \citep{Powell2019NewConstraintsDust}.  The upgrade to \texttt{CARMApy}, and the corresponding upgrade to the \texttt{ExoCARMA} base it runs on, has additionally provided greater flexibility on the types of condensates and atmospheres modelable while making the code significantly more efficient and implementing multithreading with \texttt
{OpenMP} \citep{Dagum1998OpenMPIndustryStandard}.  

This paper will describe both \texttt{CARMApy} and the underlying \texttt{ExoCARMA 2.0} version of \texttt{CARMA}.  In Section \ref{sec:history} we will describe the history of \texttt{CARMA} and document the various changes made to the software over its nearly 50 year history, in the remainder of Section \ref{sec:model} we will describe the microphysical processes, along with our modeling assumptions and numerical methods, that are used in both \texttt{CARMA} and \texttt{CARMApy}, and in Section \ref{sec:benchmark} we will demonstrate that \texttt{CARMApy} is able to replicate the results of previous versions of \texttt{CARMA} for a benchmark brown-dwarf case.  We note that the underlying code of the \texttt{CARMApy} wrapper and \texttt{ExoCARMA} is undergoing review by JOSS \citep{Cukier2026CARMApyUserFriendlyOpenSource}.

\section{Model Description}\label{sec:model}
\subsection{Code Heritage}\label{sec:history}
\begin{figure}
    \centering
    \includegraphics[width=0.95\linewidth]{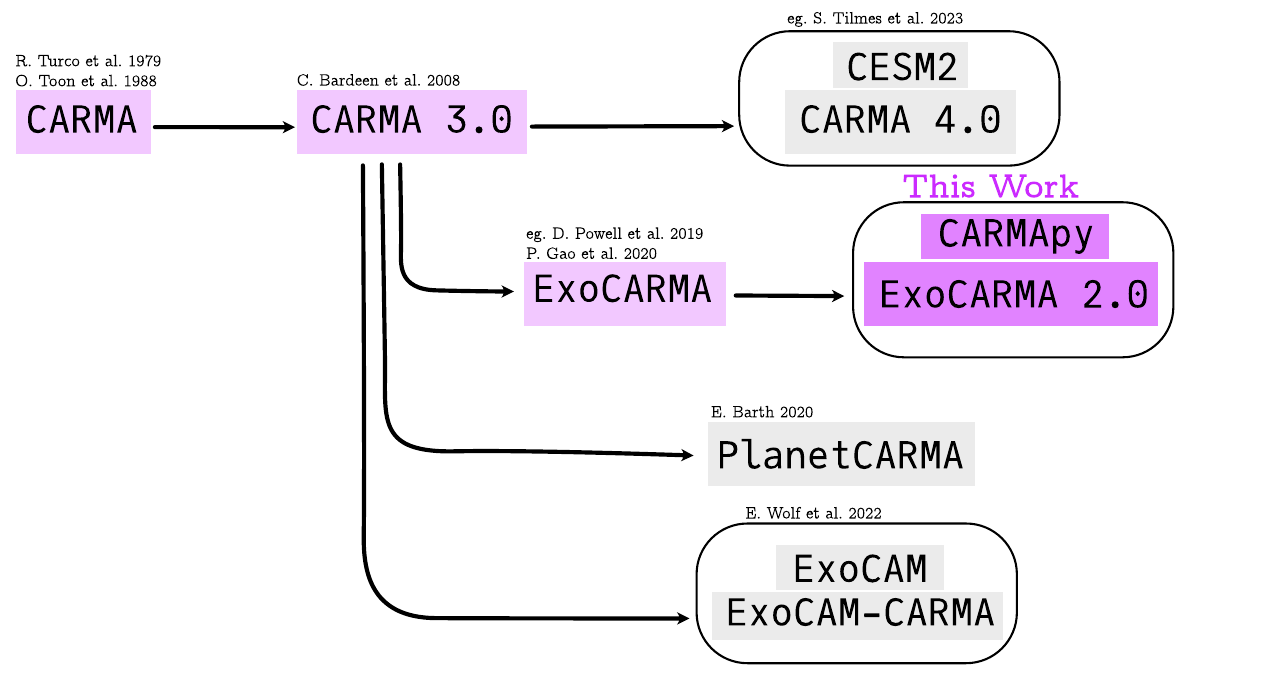}
    \caption{Diagram of the heritage of this work's version of CARMA, along with selected other versions of CARMA}
    \label{fig:map}
\end{figure}

As \texttt{CARMA} has been continuously used over almost the past 50 years, multiple versions of \texttt{CARMA} have been developed.  The original version of \texttt{CARMA} was written in \texttt{Fortran 77} by \citet{Turco1979OneDimensionalModelDescribing} and \citet{Toon1988MultidimensionalModelAerosols}.  That model was then updated to \texttt{CARMA 3.0} by \citet{Bardeen2008NumericalSimulationsThreedimensional} who modernized the code to \texttt{Fortran90}.  This main (Earth) \texttt{CARMA} branch is managed by the University Corporation for Atmospheric Research (UCAR) and is under active development today, being incorporated into the Community Earth System Model \citep[CSEM, eg. ][]{Tilmes2023DescriptionPerformanceCARMA} and is actively being used to model cloud formation on Earth \citep[eg.][]{Chen2018ImprovementStratosphericAerosol, Lian2022GlobalDistributionAsian, Rusch2017LargeIceParticles}. Additionally, branching in development history from the main \texttt{CARMA} branch are versions such as \texttt{ExoCAM-CARMA} \citep{Wolf2022ExoCAM3DClimate} which can model hazes in exoplanet atmospheres and \texttt{PlanetCARMA} \citep{Barth2020PlanetCARMANewFramework} which models clouds and hazes on solar system bodies.  Our work, \texttt{CARMApy}, is built upon the \texttt{ExoCARMA} branch of \texttt{CARMA}.  \texttt{ExoCARMA} branched from the development history of the main \texttt{CARMA} branch sometime in 2012, shortly after the release of \texttt{CARMA 3.0}.  While the main Earth \texttt{CARMA} branch has been under active development since then, the majority of updates to that main branch since \texttt{ExoCARMA} branched from it have been to the sulfate chemistry so are unlikely to be relevant to the majority of exoplanets.  We, however, intend to incorporate updates from the main Earth \texttt{CARMA} branch in the future.

This work presents two main codes---\texttt{ExoCARMA 2.0} and \texttt{CARMApy}.   \texttt{ExoCARMA 2.0} is the most up to date version of  \texttt{ExoCARMA} and includes a more flexible implementation for new condensates.  \texttt{CARMApy} is a python wrapper built on top of this latest version of \texttt{ExoCARMA}, along with a suite of convenience functions which are useful for cloud microphysical modeling.  \texttt{CARMApy} is designed in such a way that updates to \texttt{ExoCARMA} will propagate to \texttt{CARMApy}.  A diagram showing the relationship of \texttt{CARMApy} to other versions of \texttt{CARMA} is shown in Figure \ref{fig:map}.  \added{The main features added specifically in \texttt{ExoCARMA 2.0} since \texttt{ExoCARMA 1.0} are optimization including multithreading with \texttt{OpenMP}, user-definable condensates (\S \ref{sec:custom_cond}), user-definable atmospheric properties (\S \ref{sec:atmo}), as well as corrections to the sedimentation velocity (\S \ref{sec:transport}) and coagulation (\S \ref{sec:coag}) routines.  In addition to the python interface itself, the \texttt{CARMApy} wrapper also provides for helper initialization routines (\S \ref{sec:init}) as well as post-processing routines for reading results generating outputs such as spectra (\S \ref{sec:post} and \S \ref{sec:example}). }

\subsection{Cloud Microphysics}\label{sec:microphysics}
\texttt{CARMA} is a time-stepping, bin-scheme microphysical model.  This means that cloud particles can form along a size distribution independently at each pressure level with the bin resolution on this grid controlled by the user.   Upon initialization, the user specifies which microphysical processes to use among homogeneous and heterogeneous nucleation (see Sections \ref{sec:hom_nuc}, \ref{sec:het_nuc}), condensational growth / evaporation (Section \ref{sec:growth}), and coagulation (Section \ref{sec:coag}).  These processes are then simulated at each timestep, allowing particles to move from one bin to another.  Particles are also transported between altitude bins following the vertical transport and settling schema described in Section \ref{sec:transport}. 

\subsubsection{Homogeneous Nucleation} \label{sec:hom_nuc}
\texttt{CARMA} models homogeneous nucleation---the first stage of cloud formation where particles form directly from the gas phase--- following classical nucleation theory \citep{Pruppacher2010MicrophysicsCloudsPrecipitation}. Conceptually this process can be conceived of as arising out of the random motions of particles within an atmospheric cell.  The random motions of gas particles will cause the microscopic density in every given location to fluctuate as gas particles temporarily stick to each other and then fly apart.  Where these fluctuations cause a large enough \added{cluster} of stuck-together gas particles to form, these (no-longer-gas) particles become bound to each other and thus nucleation occurs.   Quantitatively, the production rate via homogeneous nucleation of particles per unit volume per unit time, $J_{\text{hom}}$ is given by \citep[Eqn. A1]{Gao2017CloudsHazesPlanetary}
\begin{equation}
    J_{\text{hom}} = 4 \pi a_c^2 \Phi Zn\exp(-\Delta F_c/kT)
\end{equation}
where $n$ is the number density of the vapor particles in the condensate's gas reservoir, $k$ is the Boltzmann constant, and $T$ is the temperature.  The critical radius, $a_c$ is defined as \citep[Eqn. 7-27]{Pruppacher2010MicrophysicsCloudsPrecipitation}
\begin{equation}\label{eqn:ac}
    a_c = \frac{2M\sigma}{\rho_p RT \ln S}
\end{equation}
where $M$, $\sigma$, $\rho_p$, and $S$ are the molar mass, surface tension, density, and saturation ratio of the condensate and $R$ is the ideal gas constant.  The energy of formation, $\Delta F_c$, of the condensate with radius $a_c$ is given by \citep[Eqns. 7-24, 7-26]{Pruppacher2010MicrophysicsCloudsPrecipitation}
\begin{equation} 
    \Delta F_c = \max_a\left(4\pi a^2\sigma - \frac{4\pi a^3}{3 M} \rho_p RT \ln S\right) = \frac{4}{3} \pi \sigma a_c^2
\end{equation}
Fundamentally, the free energy of formation is a balance between the energy due to surface tension, which grows rapidly with radius $a$, and the decrease in free energy due to the volume of the particle growing larger.  The first location where it is energetically favorable to continue growing the particle is the radius where the balance of these terms reaches its maximum, which occurs at $a=a_c$.  Following a Boltzmann distribution, there should be $n\exp(-\Delta F_c/kT)$ particles at this unstable equilibrium where just one more molecule diffusing onto them will allow them rapidly grow.  The rate at which a single molecule will diffuse onto these critically sized particles is given by $4\pi a_c^2 \Phi$ where $\Phi$ is the diffusion flux of the reservoir of gas molecules and is given by:
\begin{equation}
    \Phi = n \sqrt{\frac{kT}{2\pi m_v}}
\end{equation}
where $m_v$ is the mass of an individual gas particle.  Lastly, $Z$ is the Zeldovich factor which accounts for non-equilibrium effects, such as the fact that particles at the critical radius can have a molecule evaporate from them instead of condense as well as the deviations from equilibrium caused by the fact that nucleation inherently introduces a max flux towards larger particles, is given by (\citealt{Pruppacher2010MicrophysicsCloudsPrecipitation}, Eqn. 7-44; \citealt{Zeldovich1992TheoryNewPhase}, Eqn. 16)
\begin{equation}
    Z =  \sqrt{\frac{\Delta F''_c}{-2 \pi kT}} =\sqrt{\frac{\Delta F_c}{3 \pi k T g_m^2}}
\end{equation}
where $\Delta F''_c$ is the second derivative of the energy of formation, with respect to the number of molecule in the particle, at the critical radius, $g_m$ is the number of molecules contained in a particle of radius $a_c$.  Particles which nucleate homogeneously in \texttt{CARMA} are assigned to the particle bin \added{containing} to the critical radius, with the exact number of particles that nucleate being adjusted to conserve the total mass that nucleates.  \added{If the critical radius is smaller than the minimum bin size, the newly nucleated particles will be placed in this minimal bin, conserving mass but not particle number.}

\added{We note that the classical nucleation theory method described here carries a number of simplifying assumptions about the nucleation process such as that bulk properties including surface energy are still well defined at the nanoscales where nucleation occurs \citep{Goumans2012EfficientNucleationStardust} and that there exists exactly one relevant energy barrier for nucleation \citep{Bromley2016WhatConditionsDoes}.  Other methods for modeling nucleation exist such as non-classical nucleation theory, which uses data about the relative abundances of various cluster sizes to calculate a stationary flux through cluster size as the nucleation rate \citep{Helling2013ModellingFormationAtmospheric, Lee2015DustBrownDwarfs, Sindel2022RevisitingFundamentalProperties}. However, for these methods it is required to know the Gibbs free energies of all cluster sizes considered, derived through modeling frameworks such as density functional theory (DFT). These approaches, similar to classical nucleation theory, are therefore also fundamentally limited by our knowledge of the thermochemical properties of the nucleating species \citep{Lecoq-Molinos2024VanadiumOxideClusters} and thus require detailed DFT or similar studies so we can have a complete sample of non-classical nucleation rates for various condensates.  While updating \texttt{CARMApy} to include the ability for the user to include a cluster-data derived nucleation rate might be worthwhile, as of now only classical nucleation theory is supported.}

\subsubsection{Heterogeneous Nucleation}
 \label{sec:het_nuc}
Heterogeneous nucleation occurs when a cloud particle condensates on top of an already existing particle of a different species, known as a cloud condensation nucleus (CCN) or more informally as a ``seed particle.''  Similarly to homogeneous nucleation, heterogeneous nucleation occurs when the amount of the newly formed condensate adsorbed onto the CCN exceeds a critical amount and it becomes increasingly energetically favorable to adsorb new vapor molecules onto the heterogeneous particle.  The heterogeneous nucleation rate, in units of new particles per unit time per CCN particle is \citep[Eqn. A.6]{Gao2017CloudsHazesPlanetary}
\begin{equation}
    J_{\text{het}} = 4\pi^2r_N^2 a_c^2 \Phi c_{\text{surf}} Z_{\text{het}} \exp (-\Delta F_c f/kT), 
\end{equation} where $r_N$ is the radius of the CCN and $a_c$ is the critical radius of the cluster of newly condensed particles that form on the surface of the CCN.  Note that $a_c$ is not the value of the radius of the heterogeneous particle after nucleation.  The value of the critical radius, $a_c$, remains the same as in the homogeneous case as the local curved surface of the cluster must be in equilibrium with the vapor (once can also recover the same value for $a_c$ by calculating out the volume of a spherical cap and maximizing the free energy as in Equation \ref{eqn:ac}).  The shape factor, $f$, accounts for the fact that the surface of the cluster of newly condensed particles is not exposed to air on all sides and is given by \citep[Eqn. 9-27]{Pruppacher2010MicrophysicsCloudsPrecipitation}
\begin{equation}
    2f = 1 + \left(\frac{1 - \mu x}{\phi}\right)^4 + x^3 \left(2 - 4 f_0 + f_0^3\right) + 3\mu x^2 (f_0 - 1)
\end{equation}
where $\mu$ is the cosine of the contact angle between the condensate and the surface (with $\mu=1$ being maximally interacting and $\mu=-1$ being non-interacting) and
\begin{align}
    x &= r_N/a_c \\
    \phi &= \sqrt{1 - 2\mu x+ x^2}\\
    f_0 &= (x-\mu)/\phi
\end{align}
We assume that the primary mechanism for particle growth is diffusion along the surface of the CCN so $c_{\text{surf}}$ is the number density of condensate molecules on the surface of the CCN and is given by \citep[Eqn. A.11]{Gao2017CloudsHazesPlanetary}
\begin{equation}
c_{\text{surf}} = \frac{\Phi}{\nu} \exp(F_\text{des}/kT)
\end{equation}
Here, $\nu\exp{(-F_\text{des}/kT)}$ is the particle desorption rate, $F_\text{des}$ is the desorption energy, and the pre-exponential factor, $\nu$, can be interpreted as the attempt frequency for particle escape. We estimate $\nu$ using the energy of a harmonic oscillator ($E = 1/2 m \omega^2A^2$) following the methods of \citet[Eqn. 6]{Tielens1987CompositionStructureChemistry}\footnote{We note that this method is often cited as \citet{Hasegawa1992ModelsGasGrainChemistry} even though the authors of that work themselves cite \citet{Tielens1987CompositionStructureChemistry} for this equation}
\begin{equation}
    \nu = \sqrt{\frac{2 N_s F_\text{des}}{\pi^2m_v}} \approx \left(1.6 \times 10^{11}  \text{s}^{-1}\right)\left(\frac{F_\text{des}/k}{1 \text{K}}\right)^{1/2} M_v^{-1/2}
\end{equation}
where $M_v$ is the molar mass of the vapor particle and $N_s$ is the density of adsorption sites on the CCN and is identified with the reciprocal square amplitude of adsorbant oscillation.  $N_s$ is frequently approximated as $N_s \approx 1.5 \times 10^{15} \text{ cm}^{-2}$, which corresponds to an average distance between sites of $\sim3$\AA.  \texttt{CARMApy} defaults to assuming that the desorption energy is equal to half the latent heat of evaporation of a single molecule

Lastly, the heterogeneous Zeldovich factor is given by \citep[Eqn. 17]{Vehkamaki2007TechnicalNoteHeterogeneous}:
\begin{equation}
    Z_\text{het} = Z \sqrt{\frac{4\phi^3}{2\phi^3 + (1 - \mu x)\left(2 - 4\mu x - (\mu^2 -3)x^2\right)}}
\end{equation}

Similar to homogeneous nucleation, the $4\pi r_N^2 c_\text{surf}\exp(-\Delta F_cf/kT)$ Boltzman factor in the definition of $J_\text{het}$ can be interpreted as the number of sites on each CCN which are at the critical radius, $a_c$, and thus will continue to grow if they adsorb another particle.  Likewise the $\pi a_c^2 \Phi$ factor is the rate at which particles will reach these critical sites (note that we use a surface area of $\pi a_c^2$ instead of $4\pi a_c^2$ like we did for homogeneous nucleation because in the case of heterogeneous nucleation, the cluster of particles is touching the CCN and thus not all of the surface area is available for the adsorption of additional particles).  Note that the units of $J_\text{het}$ are in new particles per unit time per CCN, so to get the heterogeneous nucleation rate in the same units as $J_\text{hom}$ we must multiply by the number density of the CCN.  In \texttt{CARMA}, it is assumed that heterogeneous nucleation does not significantly alter the mass of the particle---as the bin sizes in \texttt{CARMA} grow exponentially larger with each successive bin, this assumption hold for most cases.
\subsubsection{Condensational Growth/Evaporation}
\label{sec:growth}
Condensation and evaporation (or more specifically, but equivalently mathematically, solid deposition and sublimation as most of our particles are solids) occur when a vapor molecule diffuses onto or away from the surface of a cloud particle.  If we assume that particles are large compared to the mean free path of the vapor ($Kn \ll 1$, we will relax this assumption later), we can invoke Fick's first law of diffusion and so the rate of change in mass, $m$, of a cloud particle can be described as \citep[Eqn. 16.1]{Jacobson2005FundamentalsAtmosphericModeling}
\begin{equation}\label{eqn:simple_growth} 
    \frac{dm}{dt} = 4 \pi R_d^2 D_v \frac{d\rho_v}{dR_d}
\end{equation}
where $R_d$ is the radial distance from the center of the particle, $d\rho_v/dR_d$ is the radial gradient of vapor density and $D_v$ is the molecular diffusion coefficient of the vapor, which is given by \citep[Eqn. 16.17]{Jacobson2005FundamentalsAtmosphericModeling}
\begin{equation} 
    D_v = \frac{5}{16 N_A d_q^2 \rho_a}\sqrt{\frac{RTM_a}{2\pi} \left(\frac{M + M_a}{M}\right)}
\end{equation}
where $M_a$ is the molar mass of the atmosphere, $N_A$ is Avogadro's number, $\rho_a$ is the local density of the atmosphere, and $d_q$ is the collision diameter of the gas molecule.

We assume that near the surface of the drop the vapor density is set by the saturation vapor pressure (along with the ideal gas law) so $\rho_v(R_d = r) = \rho_{s}$.  If we now integrate Equation \ref{eqn:simple_growth}
 from $R_d=r$ at the particles surface to $R_d=\infty$ we find that the growth/evaporation rate is given by \citep[Eqn. 16.2]{Jacobson2005FundamentalsAtmosphericModeling}
\begin{equation} 
    \frac{dm}{dt} = 4\pi r D_v(\rho_v - \rho_s)
\end{equation}
As we expect, when the atmosphere is supersaturated ($\rho_v > \rho_s$) the particles will grow and when the atmosphere is not saturated ($\rho_v < \rho_s$) the particle will evaporate.  Due to the latent heat of evaporation, however, these processes will change the temperature of the particle, thus changing their saturation vapor pressure.  To account for this, the heating due to condensation and evaporation is given by \citep[Eqn. 16.5]{Jacobson2005FundamentalsAtmosphericModeling}
\begin{equation}\label{eqn:latent_heat}
    mc_p \frac{dT_p}{dt} = L \frac{dm}{dt} - \frac{dQ}{dt}
\end{equation}
where $c_p$ is the specific heat capacity of the particle, $L$ is the latent heat of evaporation of the condensate, and $T_p$ is the temperature of the particle.  This heat is conducted away from the particle following \citep[Eqn. 16.3]{Jacobson2005FundamentalsAtmosphericModeling}
\begin{equation}\label{eqn:heat_flux}
    \frac{dQ}{dt} = -4\pi R_d^2\kappa_a \frac{dT}{dR_d}
\end{equation}
where $\kappa_a$ is the thermal conductivity of the atmosphere.  Combining Equations \ref{eqn:simple_growth}, \ref{eqn:latent_heat}, and \ref{eqn:heat_flux}, along with the ideal gas law and the Clausius-Clapeyron equation ($d \ln p_s/{dT} = L/RT^2$), and assuming the temperature of the particle is in steady state and $LM/RT - 1 \approx LM/RT$, our new equation for condensational growth / evaporation is \citep[A.15]{Gao2017CloudsHazesPlanetary}
\begin{equation} 
    \frac{dm}{dt} = \frac{4\pi r D_v p_s (S-1)}{\frac{RT}{M} + \frac{D_vL^2 Mp_s}{\kappa_a T^2}}
\end{equation}
where $p_s$ is the saturation vapor pressure.

We consider three further corrections to this equation.  Firstly, this equation was derived assuming we can treat the vapor as a continuum---this breaks down at small particle sizes where the mean free path of the vapor becomes comparable in value to the size of the particles.  To correct for this assumption we introduce a corrected diffusion coefficient and thermal conductivity \citep[Eqns. 16.19, 16.27]{Jacobson2005FundamentalsAtmosphericModeling}
\begin{align}
    D_v' &= \frac{D_v}{1 + \lambda_c \text{Kn}_c}\\
    \kappa_a' &= \frac{\kappa_a}{1 + \lambda_t \text{Kn}_t}
\end{align}
where $\lambda_c$ and $\lambda_t$ are coefficients given by \added{\citep[Eqn. 3.27]{Fuchs1971HIGHDISPERSEDAEROSOLS}}
\begin{align}
    \lambda_c &= \frac{1.33 \text{Kn}_c + 0.71}{\text{Kn}_c +1} \\
    \lambda_t &= \frac{1.33 \text{Kn}_t + 0.71}{\text{Kn}_t + 1}
\end{align}
and $\text{Kn}_c$ and $\text{Kn}_t$ are the collisional and energy exchange Knudsen numbers of the condensing gas with respect to the particle and are defined as \citep[Eqns. 16.28, 16.29]{Jacobson2005FundamentalsAtmosphericModeling}
\begin{align} 
    \text{Kn}_c &= \frac{\ell_c}{r} = \frac{3D_v}{r}\sqrt{\frac{\pi M}{8RT}} \\
    \text{Kn}_t &= \frac{\ell_t}{r} = \frac{3\kappa_a}{r\rho_a\left (c_p - \frac{R}{2M_a}\right)}\sqrt{\frac{\pi M}{8RT}} 
\end{align} where the $\ell_i$s are the mean free paths. Low Knudsen numbers indicate that the particle frequently interacts with vapor molecules and thus the continuum assumption is valid, while large Knudsen numbers imply the continuum assumption fails.  \added{We note that in the high Knudsen number limit these corrections approximate\footnote{This approximation becomes exact if we instead set the 1.33 prefactor equal to $4/3$, neglect the latent heat term (which becomes negligible in the upper atmosphere) and assume the collisional sticking coefficient is equal to one.} the collisional growth regime used in the upper atmosphere of other models \citep[i.e. ][]{Woitke2003DustBrownDwarfs}.}

Secondly, the vapor pressure over a curved surface is different than that over a flat surface---a phenomenon known as the Kelvin effect.  To correct for this effect we introduce the Kelvin factor \citep[Eqns. 16.33]{Jacobson2005FundamentalsAtmosphericModeling}
\begin{equation} 
    A_k = \frac{p_{s, \text{curved}}}{p_{s, \text{flat}}} = \exp\left(\frac{2M\sigma}{\rho_pRTr}\right)
\end{equation}
Lastly, particles which are moving generate eddies which sweep additional energy and mass onto the particle.  To account for this, we introduce two ventilation factors, $F_v$ and $F_t$, which were determined empirically from water droplets \citep[Eqns. 13-60, 13-61]{Pruppacher2010MicrophysicsCloudsPrecipitation}
\begin{align}
&&F_v &= \begin{cases} 1 + 0.108 x_v^2 &x_v \leq 1.4 \\ 0.78 + 0.308 x_v &x_v > 1.4\end{cases}   &x_v = Re^{1/2}\left(\frac{\eta_a}{\rho_aD_v'}\right)^{1/3} &&\\
&&F_t &= \begin{cases} 1 + 0.108 x_t^2 &x_t \leq 1.4 \\ 0.78 + 0.308 x_t &x_t > 1.4\end{cases}   &x_t = Re^{1/2}\left(\frac{\eta_a c_p}{\kappa'_a}\right)^{1/3}
\end{align}
where $\eta_a$ is the dynamic viscosity of the air and $Re$ is the Reynolds number (see Section \ref{sec:transport}).  After applying all of these corrections, the formula for condensational growth / evaporation used in \texttt{CARMA} is \citep[Eqn. A.16]{Gao2017CloudsHazesPlanetary}
\begin{equation}
    \frac{dm}{dt} = \frac{4\pi r D_v' p_s (S-A_k)}{\frac{RT}{MF_v} + \frac{D_v'ML^2 p_s}{\kappa_a' RT^2 F_t}}
\end{equation}

Just as with vertical transport below, \texttt{CARMA} uses the piecewise parabolic method \citep{Colella1984PiecewiseParabolicMethod} to ``transport'' particles between mass bins.

\subsubsection{Particle Sedimentation}\label{sec:transport}
The problem of particle sedimentation can be split into three regimes---laminar ($Re < 1$), transitional ($1<Re<1000$), and turbulent ($Re > 1000$).  Both the laminar and turbulent regimes have simple formulae for the sedimentation (terminal) velocities but there is no simple closed form equation for the transitional regime.  Additionally, it is required to know the velocity of particles in order to determine which fall velocity regime to use, creating a circular problem.

To resolve these issues, we begin by assuming that the particles are in the laminar regime (i.e. they are small and slow moving) and thus their sedimentation velocity, $v_t$, can be described by Stokes' fall velocity
\begin{equation} \label{eqn:stokes}
    v_t = \frac{2}{9} \frac{ g r^2 C_c \left(\rho_p-\rho_a\right)}{\eta_a} - v_\text{winds}
\end{equation}
where $g$ is the gravitational acceleration, $v_\text{winds}$ is a user defined upward wind speed, and $C_c$ is the Cunningham slip factor which is given by \citep[Eqn. 8.5]{Fuchs1964MechanicsAerosols}
\begin{equation} \label{eqn:cc}
    C_c = 1 + \text{Kn}\left(1.246  + 0.42e^{-0.87/\text{Kn}}\right)
\end{equation}
where the numerical constants in this equation are empirically derived from falling oil drops and $\text{Kn}$ is a Knudsen number given by:
\begin{equation}
    \text{Kn} = \frac{\ell}{r} = \frac{2\eta_a}{r\rho_a}\sqrt{\frac{\pi M_a}{8RT}}
\end{equation}
The Reynolds number is defined as follows
\begin{equation} \label{eqn:re}
Re = \frac{2r\rho_a(v_t +v_\text{winds} ) }{\eta_a}
\end{equation}
If $Re < 1$ then our assumption that particles are small and slow moving is valid and we use Equations \ref{eqn:stokes} and \ref{eqn:re} for $v_t$ and $Re$ respectively.  Otherwise, we must instead calculate the fall velocity in the transition ($1 < Re < 1000$) or Newtonian ($Re > 1000$) regimes.  For these regimes we introduce the drag coefficient, $C_D$, which is defined such that the terminal velocity is as \cite[Eqn. 8.44]{Seinfeld1998AtmosphericChemistryPhysics}
\begin{equation}\label{eqn:cd}
    v_t = \left(\frac{8 g r (\rho_p - \rho_a)}{3 C_D\rho_a}\right)^{1/2} - v_\text{winds}
\end{equation}
Combining this with equation \ref{eqn:re} we can define a quantity, $C_D{Re}^2$, sometimes known as the Best number, as follows \citep[Eqn. 10-142]{Pruppacher2010MicrophysicsCloudsPrecipitation}
\begin{equation}
    C_DRe^2 = \frac{32 r^3 \rho_a(\rho_p-\rho_a) g}{3 \eta^2_a}
\end{equation}
We can now use this value to find the Reynolds number using the following empirical fit \citep[Eqn. 10-145]{Pruppacher2010MicrophysicsCloudsPrecipitation}:
\begin{equation} \label{eqn:fit}
    \ln Re = B_0 + B_1x + B_2 x^2 +\ldots + B_6 x^6
\end{equation}
where $x = \ln \left(C_DRe^2\right)$ and the $B_i$ are given in Table \ref{tab:fit_coeffs}.
\begin{table}[]
    \centering
    \begin{tabular}{c|c}
    Coefficient & Value \\\hline
        $B_0$ & -3.18657 \\
        $B_1$ & +0.992696 \\
        $B_2$ & $-1.53193 \times 10^{-3}$  \\
        $B_3$ &  $-9.87059 \times 10^{-4}$ \\
        $B_4$ &  $-5.78878 \times 10^{-4}$ \\
        $B_5$ &  $+8.55176 \times 10^{-5}$ \\
        $B_6$ &  $-3.27815 \times 10^{-6}$ \\

    \end{tabular}
    \caption{Empirical coefficients for the fit given in Equation \ref{eqn:fit} \citep{Pruppacher2010MicrophysicsCloudsPrecipitation}.}
    \label{tab:fit_coeffs}
\end{table}
If $Re < 1000$ we then calculate the fall velocity from Equation \ref{eqn:re} as follows 
\begin{equation}
    v_t = \frac{\eta_a Re}{2 \rho_a r} - v_\text{winds}
\end{equation}
If instead $Re > 1000$, we set $C_D = 0.45$ and calculate the fall velocity from Equation \ref{eqn:cd}.  

\subsubsection{Eddy Diffusion}
In addition to particle sedimentation, \texttt{CARMA} also tracks particle diffusion, allowing for the vertical lofting of cloud particles.  We assume that eddy (turbulent) diffusion dominates molecular diffusion, as is expected to be true in all but the very upper atmosphere \cite{Woitke2020DustBrownDwarfs}, and thus the diffusive velocities are given by
\begin{equation}
    v_{ed} = -K_{zz} \frac{d\ln f_r}{dz}
\end{equation}
where $f_r$ is the particle mixing ratio and where $K_{zz}$ is the eddy diffusion coefficient.

As in for particle advection in mass space, particles are transported between altitude bins using the piecewise parabolic method.

\subsubsection{Coagulation/Coalescence}
 \label{sec:coag}
When two condensate molecules collide with each other, depending on their mass and momenta, there are largely three classes of outcomes---they can stick together, they can bounce off each other, or they can fragment into pieces \citep{Guttler2010OutcomeProtoplanetaryDust}.  To model these collisional processes, we thus must model both the rate at which particles collide and then the outcomes of that collision.  \texttt{CARMA} considers three sources for collisions----Brownian motion, convective diffusion enhancement, and gravitational coalescence (also known as differential settling).  Brownian motion and gravitational coalescence are common processes considered in other microphysical cloud models \citep{Ohno2017CondensationcoalescenceCloudModel, Lee2025MonoculturePolydisperseMoment}, whereas convective diffusion enhancement serves as a correction term for the Brownian motion term.  We note that while others have found that large scale atmospheric turbulence can be an important source of collisions near the cloud base \citep{Samra2022MineralSnowflakesExoplanets},   \texttt{CARMA} does not currently consider turbulence as a collision source.  \texttt{CARMA} assumes that Brownian motion (and thus convective diffusion enhancement) causes particles which collide to stick together and form compact, spherical (ie non-fractal) particles with 100\% probability while \texttt{CARMA} allows particles that collide due to differential settling to either bounce or stick according to a formalism discussed below.  \texttt{CARMA} does not consider fragmentation as an outcome of collisions.

The Brownian coagulation kernel, for coagulation between two particles, 1 and 2, $K^b_{12}$, which represents the collision probability of two particles undergoing thermal motion is given by
(\citealt{Fuchs1964MechanicsAerosols}, Eqns. 49.26, \citealt{Jacobson2005FundamentalsAtmosphericModeling}, Eqns. 15.33)
\begin{equation}
    K^B_{12} =  K^B_{21} = 4 \pi (D_1 + D_2)(r_1 + r_2) \beta 
\end{equation}
Where the $D_i$ and $r_i$ are the molecular diffusion coefficients and radii of the populations of particles and $\beta$ is a interpolation factor that interpolates between the continuum and kinetic regimes.  The molecular diffusion coefficients are given by \citep[Eqn. 15.29]{Jacobson2005FundamentalsAtmosphericModeling}
\begin{equation}
    D_i = \frac{kTC_c}{6\pi\eta_ar_i}
\end{equation}
where $C_c$ is the Cunningham slip factor (see Equation \ref{eqn:cc}).  The interpolation factor, $\beta$, is given by \citep[Eqn. 49.27]{Fuchs1964MechanicsAerosols}
\begin{equation}
    \frac{1}{\beta} = \frac{r_1 + r_2}{r_1 + r_2 + \sqrt{\delta_1^2 + \delta_2^2}} + \frac{4(D_1 + D_2)}{(r_1+r_2)\sqrt{v_1^2 + v_2^2}}
\end{equation}
where the $v_i$ are the thermal velocities of the particles and are given by
\begin{equation}
    v_i = \sqrt{\frac{8kT}{\pi m_i}}
\end{equation}
where $m_i$ is the mass of the particles.  The transition length scale, $\delta_i$, is given by \citep[Eqn 15.34]{Jacobson2005FundamentalsAtmosphericModeling}
\begin{equation}
    \delta_i = \frac{(2r_i + \ell_i)^3 - (4r_i^2 + \ell_i^2)^{3/2}}{6 r_i \ell_i} -2r_i
\end{equation}
with $\ell_i$ being the mean free path of the particle given by
\begin{equation}
    \ell_i = \frac{8D_i}{\pi v_i}
\end{equation}

In a similar mechanism to the ventilation coefficients discussed in Section \ref{sec:growth}, falling large particles generate eddies which can draw smaller particles onto them.  This convective diffusion enhancement increases the coagulation rate of large particles and its kernel, $K^{DE}$, is given by \citep[Eqn. 15.35]{Jacobson2005FundamentalsAtmosphericModeling}
\begin{equation}
    K^{DE}_{12} = \begin{cases} \alpha_{lo} K^{B}_{12} Re_j^{1/3} \left(\frac{\eta}{\rho D_i}\right)^{1/3} & Re_j \leq1  \\
    \alpha_{hi}K^{B}_{12} Re_j^{1/2} \left(\frac{\eta}{\rho D_i}\right)^{1/3} & Re_j \geq1
    \end{cases},\qquad (r_j \geq r_i)
\end{equation}
where we take $\alpha_{lo} = \alpha_{hi} = 0.45$.\footnote{See Chapter 17 of \citet{Pruppacher2010MicrophysicsCloudsPrecipitation} for discussion of this value.} 

Lastly, we consider the effects of gravitational coalescence.  As particles fall with differing fall velocities they can collide and then coalesce together.  The kernel for this process is given by
\begin{equation}
    K^{G}_{12} = E_\text{collide} E_\text{coalesce} \pi (r_1 + r_2)^2 \left|v_{t,1} - v_{t,2}\right|
\end{equation}
where $E_\text{collide}$ and $E_\text{coalesce}$ are the collision and coalescence efficiencies respectively and $v_{t,1}$ and $v_{t,2}$ are the terminal fall velocities of the two particles.

We set $E_\text{collide} = \max(E_\text{Langmuir}, E_\text{Fuchs})$  
 where $E_\text{Langmuir}$ is given by \citep[Eqn. 23]{Langmuir1948PRODUCTIONRAINCHAIN}
 \begin{equation}
          E_\text{Langmuir} = \frac{ E_V + E_A Re_j/60}{1 + Re_j/60} \qquad (r_j \geq r_i)
 \end{equation} where $E_V$ and $E_A$ are the efficiencies in the viscous ($Re_j \ll 1$) and aerodynamic ($Re_j \gg 1$) cases respectively and are given by \citep[Eqns. 16, 23]{Langmuir1948PRODUCTIONRAINCHAIN}
 \begin{align}
     E_V &=\begin{cases}
         \left(1 + \frac{0.75 \ln 2 S_k}{S_k - 1.214}\right)^{-2} & S_k > 1.214 \\
         0 & S_k \leq 1.214
     \end{cases}\\
     E_A &= \begin{cases}
         \frac{S_k^2}{(S_k + 1/2)^2}\hspace{1.5em} & S_k \geq 1/12 \\
         0& S_k < 1/12
     \end{cases}
 \end{align}
 where $S_k$ is the Stokes number and is given by
 \begin{equation}
     S_k = \frac{v_{t,i} |v_{t, j} - v_{t,i}|}{r_j g},\qquad (r_j \geq r_i)
 \end{equation}
The $E_\text{Fuchs}$ term in our definition of $E_\text{collide}$ is the collision efficiency given by  \citep[Eqn. 11-86]{Fuchs1951TheoryOverheadPrecipitation, Friedlander1957MassHeatTransfer, Pruppacher2010MicrophysicsCloudsPrecipitation}, designed to correct $E_\text{Langmuir}$ for the direct interception caused by non-point fine particles \citep[see][\S 54]{Fuchs1964MechanicsAerosols}
\begin{equation}
    E_\text{Fuchs} = \frac{(r_i/r_j)^2}{2\left(1+ (r_i/r_j)\right)^2}, \qquad r_j > r_i
\end{equation}

 Finally, $E_\text{coalesce}$, which captures the effect that particles may collide and then `bounce-off,' is empirically determined for raindrops and is given by \citep[Eqn. 4]{Beard1984CollectionCoalescenceEfficiencies}:
 \begin{equation}
    E'_\text{coalesce} =  (a-b)^{1/3} - (a+b)^{1/3} + 0.459
 \end{equation}
where
\begin{align}
    a &=\sqrt{b^2 + 0.00441}\\
    b &= 0.0946\beta - 0.319 \\
    \beta &= \ln\left(\frac{r_i}{1\text{ }\mu\text{m}}\right) + 0.44 \ln\left(\frac{r_j}{200\text{ }\mu\text{m}}\right),\qquad(r_j > r_i)
\end{align}
where we bound $E_\text{coalesce}$ to be between 0.5 and 1.0; that is
\begin{equation}
    E_\text{coalesce} = \max(0.5, \min(1.0, E'_\text{coalesce}))
\end{equation}
The total coagulation kernel is now given by:
\begin{equation}
    K^{tot}_{12} = K^B_{12} + K^{DE}_{12} + K^G_{12}
\end{equation}

\subsubsection{A Note on Supersaturation Ratios}

The standard supersaturation ratio of a vapor, $S$, is defined as the ratio of the partial pressure of the vapor $p_v$ to the saturation vapor pressure, $p_s$.  That is:
\begin{equation}
    S = \frac{p_v}{p_s}
\end{equation}

For Type III reactions \cite[][Appendix B]{Helling2006DustBrownDwarfs}, however,  which involve more than one molecular product or more than one molecular reactant, grain chemistry models show that the effective supersaturation ratio for the reaction can instead be expressed as
\begin{equation}
    S_r = S^{1/\nu_r^\text{key}}
\end{equation}

where, $\nu_r^\text{key}$, is the stoichiometric ratio of the limiting reactant in the reaction.  For the above microphysical equations, this effective supersaturation ratio, $S_r$, is used in place of the standard supersaturation ratio for type III reactions.

\subsection{Numerical Methods}

Given the number density vector of all gas and particle bins at timestep $t$, $\mathbf{n}^t$, the number density vector at the next time step can be determined by simultaneous application of operators, $\mathcal{L}_\text{proc}$, representing the effects of the various modeled processes:
\begin{equation}
    \mathbf{n}^{t+1} =   (\mathcal{L}_\text{transport} + \mathcal{L}_\text{coagulation} + \mathcal{L}_\text{growth/evap} + \mathcal{L}_\text{nuc} + \mathcal{L}_\text{gas depletion}) \mathbf{n}^t
\end{equation}
  \texttt{CARMA} implements an operator splitting method where in each timestep vertical transport is first calculated; then coagulation; then nucleation, growth, and evaporation; and then finally gas depletion.  Put formally in our operator notation that is
\begin{equation}
    \mathbf{n}^{t+1} =   \mathcal{L}_\text{gas depletion}(\mathcal{L}_\text{growth/evap} + \mathcal{L}_\text{nuc})  \mathcal{L}_\text{coag}\mathcal{L}_\text{transport}\mathbf{n}^t
\end{equation}
The following subsections describe how \texttt{CARMA} implements these various operators.

\subsubsection{Particle Transport}
Cloud particles and gas are advected between bins (both altitude level bins, and for clouds, particle mass bins) using the piecewise parabolic method (PPM).  A full description of this method is presented in Section 1 of \citet{Colella1984PiecewiseParabolicMethod}---we only outline the broad steps of the method.  The PPM is used to solve the linear advection equation
\begin{equation}\label{eqn:adv}
    \frac{\partial n}{\partial t} + u\frac{\partial n}{\partial \xi} = 0
\end{equation}
where $n$ is the particle concentration, $\xi$ is a generalized coordinate (eg. altitude or mass), and $u$ is a generalized advection velocity.  If we let $n_k$ be the particle concentration in bin $k$ and let $\xi_{k-1/2}$ and $\xi_{k+1/2}$ be the boundaries of that bin, the relationship between the continuous particle distribution at timestep $t$, $n(\xi, t)$, and the binned particle concentration at the same timestep is simply the averaged concentration across the bin:
\begin{equation}
    n_k^t = \frac{1}{\xi_{k+1/2} - \xi_{k-1/2}}\int_{\xi_{k-1/2}}^{\xi_{k+1/2}} n(\xi, t)d\xi
\end{equation}
Similarly, we can update this distribution by solving Equation \ref{eqn:adv} as follows:
\begin{equation}
    n_k^{t+1} = \frac{1}{\xi_{k+1/2} - \xi_{k-1/2}}\int_{\xi_{k-1/2}}^{\xi_{k+1/2}} n(\xi - u\Delta t, t)d\xi
\end{equation}
The problem with this method, however, is we do not know the continuous distribution $n(\xi, t)$, we only know the discrete distribution $n_k^{t}$.  The PPM scheme solves this problem by introducing parabolic interpolation functions in each bin.  These parabolas are chosen such that existing local extrema are preserved and no local extrema are introduced.  These conditions require that the parabolae are monotonic within each bin, and constant at the local minima and maxima of the discrete distribution.  

We seek a form for the advective flux across each boundary which splits into the upwards and downwards flux where each term only depends on the number density in one bin (dropping the $t$ for clarity):

\begin{equation}
    F_{\text{ad, }k+1/2} = n_{k} v^{\uparrow}_{ \text{ad, } k+1/2}  - n_{k+1} v^{\downarrow}_{ \text{ad, } k+1/2} 
\end{equation}



If at the $k+1/2$ boundary $u_{k+1/2} > 0$, then $v^{\downarrow}_{\text{ad, }k+1/2} = 0$ and
\begin{equation}
    F_{\text{ad, }k+1/2} = n_{k} v^{\uparrow}_{\text{ad, }k+1/2}
      = \frac{1}{\Delta t}\int_{\xi_{k+1/2} - u_{k+1/2}\Delta t}^{\xi_{k+1/2}} n(\xi)\,d\xi
\end{equation}
Similarly, if $u_{k+1/2} < 0$ then $v^{\uparrow}_{\text{ad, }k+1/2} = 0$ and
\begin{equation}
    F_{\text{ad, }k+1/2} = -n_{k+1} v^{\downarrow}_{\text{ad, }k+1/2}
      = -\frac{1}{\Delta t}\int_{\xi_{k+1/2}}^{\xi_{k+1/2} - u_{k+1/2}\Delta t} n(\xi)\,d\xi
\end{equation}

\subsubsection{Eddy Diffusion}
The net flux due to diffusion, $F_\text{dif}$, can be described as
\begin{equation}
    F_\text{dif} = nv_\text{ed} = -n K_{zz} \frac{d\ln f_r}{dz} = -\rho_a K_{zz} \frac{df_{r}}{dz}
\end{equation}
We similarly seek an expression for the flux across the bin boundaries of the form
\begin{equation}
    F_{\text{dif, } k+1/2} = n_{k} v^{\uparrow}_{ \text{dif, } k+1/2}  - n_{k+1} v^{\downarrow}_{\text{dif, } k+1/2}
\end{equation}
Assuming that between bins the atmospheric pressure profile follows an exponential profile, the effective velocity coefficients can be written as
\begin{align}
    v^{\uparrow}_{\text{dif, } k+1/2} &= \frac{K_{zz}}{\Delta z} \frac{\ln \left(\rho_{k}/\rho_{k+1}\right)}{\rho_{k}/\rho_{k+1} -1}\\
    v^{\downarrow}_{\text{dif, } k+1/2} &= \frac{\rho_k}{\rho_{k+1}} v^{\uparrow}_{ \text{dif, } k+1/2}
\end{align}
Now having the effective velocities for each of these processes at each boundary between bins, \texttt{CARMA} solves the following mass conservation equation
\begin{equation}
    \frac{\partial n}{\partial t} + \nabla \cdot (F_\text{dif} + F_\text{ad}) = 0
\end{equation}

For a given layer we can write this equation as
\begin{equation}
     n_k^{t+1} - n_k^t = \frac{\Delta t}{\Delta z} \left(v^\uparrow_{k-1/2} n_{k-1}^* + v ^\downarrow_{k+1/2}n_{k+1}^* - \left(v^\downarrow_{k-1/2} + v^\uparrow_{k+1/2}\right)n_k^*\right)
\end{equation}
where $v^\uparrow_{k+1/2} = v^\uparrow_{\text{ad, }k+1/2} + v^\uparrow_{\text{dif, }k+1/2}$ and so on. We use the notation $n^*$ on the right hand side in this equation as \texttt{CARMA} can either solve this equation explicitly, using $n^t$ in place of $n^*$, or implicitly, using $n^{t+1}$ instead.  For efficiency purposes, if the entire column is stable under the explicit scheme ($\frac{\Delta z_k}{\Delta t} > v^\downarrow_{k-1/2} + v^\uparrow_{k+1/2}$
for all $k$), the explicit scheme is used for the column.  Otherwise \texttt{CARMA} switches to the implicit scheme and uses the Thomas tridiagonal Algorithm \citep[][\S A.2]{Durran1999NumericalMethodsWave} as the transport equation describes a tridiagonal matrix which is diagonally dominant, so the algorithm is guaranteed to be stable.

\subsubsection{Coagulation}

Following \citet{Jacobson1994ModelingCoagulationParticles}, the changes in particle abundance due to coagulation is calculated using a semi-implicit method.  The equation which we solve is
\begin{equation}
    \frac{dn}{dt} = S - Ln
\end{equation}
where $S$ is the source (production) term and $L$ is the loss rate.  Put in semi-implicit finite difference form, this is
\begin{equation}
    \frac{n_k^{t+1} - n_k^t}{\Delta t} = S_k - L_k n_k^{t+1}
\end{equation}
with $S$ and $L$ calculated at the start of the timestep.  This means that the updated number density is thus
\begin{equation}
    n_k^{t+1} = \frac{n_k^t + S_k\Delta t}{1 + L_k \Delta t}
\end{equation}
This implicit form guarantees that particle concentration never becomes negative.  

The loss term, assuming the resulting product completely leaves bin $k$, for example by colliding with a different species, is
\begin{equation}
    L_k = \sum_{j} K_{jk} n_j^t
\end{equation}

Under the same assumption, the source term can be written as\footnote{To clarify the notation: $i<j$ if and only if $m_i < m_j$ and $i$ and $j$ are of the same condensate species, even if different sizes}
\begin{equation}
    S_k = \sum_{k < i+j < k+1}K_{ij} n_i^{t+1} n_j^t P_{ijk}^\downarrow + \sum_{k-1 < i + j < k}K_{ij} n_i^{t+1} n_j^t P_{ijk-1}^\uparrow
\end{equation}

where $P_{ijk}^\downarrow$ and $P_{ijk}^\uparrow$ are partition factors to account for the fact that fact that the mass of created particle likely falls between bins. Since the sum treats the orderings $(i,j)$ and $(j,i)$ as separate terms,  $P_{ijk}^\downarrow$ and $P_{ijk}^\uparrow$ can be straightforwardly derived from the conservation of the source particle's mass ($P_{ijk}^\downarrow m_k +P_{ijk}^\uparrow m_{k+1} = m_i$) and normalization to the mass fraction ($P_{ijk}^\downarrow + P_{ijk}^\uparrow = m_i/m_{i+j}$):
\begin{align}
   P_{ijk}^\downarrow &=  \frac{m_i}{m_{i+j}}\frac{r_m m_k - m_{i+j}}{(r_m-1) m_k}  \\
    P_{ijk}^\uparrow &=  \frac{m_i}{m_{i+j}}\frac{m_{i+j} - m_k}{(r_m-1) m_k} 
\end{align} 
where $r_m = m_{k+1}/m_k$ is the mass ratio between bins and $m_{i+j} = m_i + m_j$ (the $P_{ijk}$ are bounded between 0 and 1). Note that $P^\uparrow_{ijk} + P^\uparrow_{jik} + P^\downarrow_{ijk}+ P^\downarrow_{jik} = 1$ which is the expected number conservation. Also note that in our calculation of $S_k$ we consider the number density of the source ($n_i^{t+1}$) and partner ($n_j^t$) species at different time steps.  This is because if we were to make this term fully implicit with the partner species being evaluated also at $t+1$ we would break mass conservation---the loss term is calculated semi-implicitly so the production term also needs to be calculated semi-implicitly.

In cases of self collision where part of the resulting particle lies still within the source bin ($m_i < m_{i+j} < m_{i+1}$), which might happen if a particle collides with a particle in the same bin or with a smaller particle of the same species, we consider the part of the source particle that remains to contribute neither to the production nor the loss.  This means that we set $P^\downarrow_{ijk} = 0$ if $i=k$ (note that the $P^\uparrow_{ijk}$ remains as is as it is used for the updating bin $i+1$, not bin $i$).  We similarly adjust the loss term to account for the fact that we do not treat the $P^\downarrow_{ijk}$ as part of the production term when $i=k$:
\begin{equation}
    L_k = \sum_{j+k \nless k+1}K_{kj}n_j^t + \sum_{j+k < k+1} K_{kj}n_j^t \left(1 - P^\downarrow_{kjk}\right)
\end{equation}

\subsubsection{Growth, Evaporation, and Nucleation}

Similar to coagulation, growth, evaporation, and nucleation are simultaneously calculated using a semi-implicit method as follows
\begin{equation}
    n_k^{t+1} = \frac{n_k^t + S_k\Delta t}{1 + L_k \Delta t}
\end{equation}

The source term for these processes can be written as
\begin{equation}\label{eqn:hetevap}
    S_k= J_\text{hom} + \sum_i J_{ki}n_i^t +v^\uparrow_{\text{grow, } k-1/2} n^t_{k-1} +v^\downarrow_{\text{evap, } k+1/2} n_{k+1}^t + S_\text{het evap}
\end{equation}
where $\sum_i J_{ki}n_i^j$ describes all the heterogeneous nucleation events which produce particles in bin $k$ where particles in bin $i$ served as CCN,  $v^\uparrow_{\text{grow, } k-1/2}$ and $v^\downarrow_{\text{evap, } k+1/2}$ are analogous to $v^\uparrow_{\text{ad, } k-1/2}$ and $v^\downarrow_{\text{ad, } k+1/2}$ and describe the growth rate from the bin below and evaporation rate from the bin above respectively, and $S_\text{het evap}$ describes the cores left behind by evaporating heterogeneous species.

The loss term can be written as
\begin{equation}
    L_k = \sum_i J_{ik} + v^\uparrow_{\text{grow, } k+1/2}  +v^\downarrow_{\text{evap, } k-1/2}
\end{equation}
where $\sum_i J_{ik}$ describes the nucleation events where the current bin, bin $k$, serves as the CCN.

\subsubsection{Gas Depletion}
After the rest of the processes are considered, \texttt{CARMA} depletes the gas phase in an explicit forward Euler scheme conserving mass from timestep to timestep.  That is \texttt{CARMA} calculates the amount of new condensate created or destroyed in each layer by nucleation, growth, and evaporation and then adjusts the gas phase to balance those processes.  Numerically this means that
\begin{equation}
    n_\text{gas}^{t+1} - n_\text{gas}^t= \sum_\text{cloud} \left(n_\text{cloud}^{t+1} - n_\text{cloud}^{t}\right) \frac{m_\text{cloud}}{m_\text{gas}}\frac{\nu_\text{cloud}^\text{gas}M_\text{gas}}{M_\text{cloud}}
\end{equation}
where the sum is over every bin of every cloud species and $\nu_\text{cloud}^\text{gas}$ is the stoichiometric ratio of the reaction that forms the condensate species for the gas in question.  As the gas depletion scheme in \texttt{CARMA} is explicit it can cause the gas abundance to become negative.  If this is the case, \texttt{CARMA} substeps at half the timestep, restarting the growth, evaporation, and nucleation calculations.

\added{\subsubsection{Time Stepping}}
\added{In \texttt{CARMA}, the length of a single timestep is chosen by the user.  As \texttt{CARMA} uses explicit Euler methods in addition to implicit Euler methods for advancing the simulation state by a timestep, timesteps which are chosen to be too large may cause the code to diverge from the true solution.  \texttt{CARMA} attempts to detect this by checking for unphysical conditions after each timestep such as negative gas abundances and will restart the timestep substepping with half the timestep in these cases.  Care must still be taken when picking a timestep to not pick a timestep which is too large as other unphysical simulation states, such as a detached lower cloud deck in an atmosphere without an inversion, may still occur when the timestep is too large.}

\added{\subsubsection{Boundary Conditions}}
\added{\texttt{CARMA} allows for three different boundary condition types: "Fixed Concentration," "Fixed Flux," and "Zero Grad." These boundary conditions are set at both the top and bottom of the atmosphere and separately for both particle and gas species.  Note that the boundary condition type must be the same for all gas species, and separately the same for all particle species, although the specific flux on concentration chosen may differ across different gas and particle species.  The "Fixed Concentration" boundary condition sets the concentration of the species at the boundary to a user specified value, the "Fixed Flux" boundary condition sets the flux of the species across the boundary to the user specified value, and the "Zero Grad" models the species with zero concentration gradient reducing diffusive fluxes to only those driven by density difference while still allowing advective fluxes caused by phenomena such as sedimentation winds to transport material across the boundary.}

\added{The default behavior in \texttt{CARMApy} is to set the top of atmosphere for both particles and gas to have a flux of 0, and for the bottom of the atmosphere to have the gas fixed in concentration at the user or \texttt{FastChem} (see Section \ref{sec:fastchem}) set reservoir value and for the particle concentration to default to 0.}

\subsubsection{A Note on Heterogeneous Particles}
If a particle is heterogeneous, meaning it formed by heterogeneous nucleation or coagulation between two particles of different compositions, \texttt{CARMA} treats this particle as a shell material which is exposed to the gas phase and a core material which is isolated from the gas phase.  This means that when these particles grow or evaporate, they use only the shell material as the material properties for these processes.  Additionally, as these particles grow, evaporate, coagulate, and move around in the atmosphere, \texttt{CARMA} tracks the total amount of mass that is trapped in the core in each size and altitude bin.  If at any point, the average mass in the core exceeds the mass of the bin, the particles are considered to have evaporated their shell material. The core material is then deposited into the pure material's bins into the two bins closest to the heterogeneous species mass bin, conserving both mass and number. This process uses partition functions analogous to those used in coagulation and determines the $S_\text{het evap}$ term in Eqn. \ref{eqn:hetevap}.  


\subsection{Atmospheric Characteristics}\label{sec:atmo}
Some of the above microphysical processes depend on either the thermal conductivity of the atmosphere, the viscosity of the atmosphere, and/or the specific heat capacity of the atmosphere.  \texttt{CARMApy} requires a constant heat capacity but allows for the thermal conductivity and viscosity to change as a function of temperature.  The thermal conductivity of the atmosphere, $\kappa_a$ is parameterized as follows:
\begin{equation}
    \kappa_a = \kappa_0 + \kappa_1T + \kappa_2T^2
\end{equation}
with the $\kappa_i$ being user specified parameters.  The dynamic viscosity of the atmosphere, $\eta_a$, is parameterized as follows
\begin{equation}
    \eta_a = \frac{\eta_1\left(T/ 1\text{ K}\right)^{\eta_2}}{1 + \eta_3/T + \eta_4/T^2}
\end{equation}
where the $\eta_i$ are parameters fit to empirical data.

\texttt{CARMApy} includes a default parameter set assuming a pure H$_2$ atmosphere, which is a reasonable assumption for bodies such as hot Jupiters and brown dwarfs, and allows users to set the atmospheric parameters to user defined values.  The \texttt{CARMApy} default atmospheric parameters are presented in Table \ref{tab:atmos}.  We point the reader to \citet{Lemmon2025ThermophysicalPropertiesFluids} for thermal conductivity values and to Table 2-138 of \citet{Green2019PerrysChemicalEngineers} for viscosity values of alternate background gases.  We plan that future versions of \texttt{CARMApy} will include a greater selection of default values as well as schemes for interpolating between them.

\begin{table}[]
    \centering
    \begin{tabular}{c|cc}
        Symbol & Value & Reference \\\hline
        $c_p$ & $1.3 \times 10^8$ erg/g/K & \citet{Kataria2015AtmosphericCirculationHot} \\
        $\kappa_0$ & 7992.77 ergs/s/cm/K & Fit to \citet{Lemmon2025ThermophysicalPropertiesFluids} \\ 
        $\kappa_1$ & $38.08$ ergs/s/cm/K$^2$ &  \\
        $\kappa_2$ & $-1.2585 \times 10^{-4}$ ergs/s/cm/K$^3$ &  \\
        $\eta_1$ & $1.7970\times 10^{-6}$ poise &  \citet{Green2019PerrysChemicalEngineers}, Table 2-138 \\ 
        $\eta_2$ & 0.685 &   \\
        $\eta_3$ & -0.59 K &   \\
        $\eta_4$ & 140 K$^{2}$ &   

    \end{tabular}
    \caption{\texttt{CARMApy} default parameters for a pure H$_2$ atmosphere.}
    \label{tab:atmos}
\end{table}

\subsection{\texttt{CARMApy} Condensates}\label{sec:custom_cond}
\texttt{CARMApy} allows for the modeling of most if not any condensate assuming its material properties are known.  To specify a condensate, the density, molar mass, collisional diameter, saturation vapor pressure function, and surface tension function must be known.  Each condensate is assumed to condense from a single limiting gas and the molar mass of that gas, the stoichiometry factor of the condensation reaction, and whether or not the condensation reaction is considered a type III reaction (see \citet{Helling2006DustBrownDwarfs}) must also be specified.  Lastly, if the condensate is formed through heterogeneous nucleation then the cosine of the contact angle between the gas species and the seed particle for the nucleation reaction must be specified. 

It is assumed that the saturation vapor pressure function takes the following form:

\begin{equation}
    \log_{10} \frac{p_s}{(10^6 \text{ barye})} = \alpha_0 - \frac{\alpha_1}{T} - \alpha_2 [\text{Fe}/\text{H}] - \alpha_3\log_{10} \frac{P}{(10^6 \text{ barye})}
\end{equation}
where $T$ is the temperature, $P$ is the atmospheric pressure, and the $\alpha_i$ are coefficients unique to each condensate.  Similarly it is assumed the surface tension of the air-condensate boundary, $\sigma$ takes the following form:

\begin{equation}
\sigma = \sigma_0 - \sigma_1 T
\end{equation}
where as before, $\sigma_0$ and $\sigma_1$ are coefficients unique to each condensate.

While still possible to specify, \texttt{CARMApy} defaults to calculating the latent heat of vaporization, $L$, using the saturation vapor pressures and the Clausius-Clapeyron equation:
\begin{equation}
    \frac{d \ln p_s}{dT} = \frac{L}{RT^2}
\end{equation}
where $R$ is the ideal gas constant.

Lastly, for heterogeneously nucleating species, the contact angle, $\mu$, can be determined through Young's relation:
\begin{equation}
    \mu = \frac{\sigma_\text{core} - \sigma_\text{interface}}{\sigma_\text{cond}}
\end{equation}
where $\sigma_\text{cond}$ is the surface tension of the condensate, $\sigma_\text{core}$ is the surface tension of the core, and $\sigma_\text{interface}$ is the surface tension of the core-condensate boundary.  We most often assume $\sigma_\text{interface} = 0$ due to a lack of laboratory measurements.

The values for all the constants assumed for the \texttt{CARMApy} default condensates are given in Appendix \ref{sec:defaults}.  These constants come from a combination of standard reference databases, density functional theory and molecular dynamics simulations, thermodynamical modeling, and lab experiments.  While some of these values are well constrained, a number of them rely on simplifying modeling assumption and/or extrapolation to temperature and pressure regimes far beyond where they were originally contained.  \added{For further discussion of which species are likely to condense and/or serve as seed particles, as well as their properties, we point the reader to \citet{Visscher2006AtmosphericChemistryGiant}, \citet{Visscher2010ATMOSPHERICCHEMISTRYGIANT}, \citet{Morley2012NeglectedCloudsDwarf}, and \citet{Lee2018DustBrownDwarfs}.  In addition to the previous references, extensive lists of condenstate forming reactions can be found in \citet{Helling2008DustBrownDwarfs} and \citet{Helling2019SparklingNightsVery}.  For work using DFT and quantum chemical calculations to constrain the nucleation properties of particular species we point the reader to \citet{Lee2015DustBrownDwarfs}, \citet{Sindel2022RevisitingFundamentalProperties}, \citet{Lecoq-Molinos2024VanadiumOxideClusters}, and \citet{Bisht2026MetalOxideClusters}.}  Further work along these lines constraining the material properties of cloud particles in regimes relevant to exoplanet and brown dwarfs contexts is needed to tamp down on uncertainties in our models.

\subsection{Optional Model Initialization Routines}\label{sec:init}
\subsubsection{Cloud Base Chemical Equilibrium} \label{sec:fastchem}
\texttt{CARMA} does not include any chemistry calculations other than the cloud microphysics discussed above.  This means that a gas will remain at a constant mixing ratio in \texttt{CARMA} in regions that do not contain any cloud particles.  Thus, to ensure that the cloud base has an accurate mixing ratio of the gas phase, it is recommended to set the bottom boundary condition for the gas to be a fixed concentration equal to the concentration at the cloud base---\texttt{CARMApy} contains convenience features\footnote{see \href{https://carmapy.readthedocs.io/en/latest/_autosummary/carmapy.chemistry.html\#carmapy.chemistry.populate_abundances_at_cloud_base}{\texttt{carmapy.chemistry.populate\_abundances\_at\_cloud\_base()}} } that allow the user to easily do precisely this.

After calculating the saturation vapor pressure for the relevant gas species across the provided $T$-$P$ profile, $p'_i$, \texttt{CARMApy} uses \texttt{pyFastChem} \citep{Stock2022FASTCHEM2Improved} to calculate the equilibrium partial pressures, $p_i$ of the gas across those same points.  For this process, the cloud base is assumed to form at the location given by equilibrium cloud condensation theory, that is where $p_i' = p_i$.  The mixing ratio of the gas at this location (calculated by dividing the partial pressure by the atmospheric pressure) is then set as the boundary condition at the bottom of the atmosphere for the gas.

\subsubsection{Adiabatic Atmospheric Extension}
It is occasionally useful to extend the atmospheric profile given to higher pressures in order to capture the entirety of the cloud formation process for which we provide a convenience function\footnote{see \href{https://carmapy.readthedocs.io/en/latest/_autosummary/carmapy.Carma.html\#carmapy.Carma.extend_atmosphere}{ \texttt{carmapy.Carma.extend\_atmosphere()}}}.  To this end, we assume the deep atmosphere is a convective zone and thus follows an adiabatic temperature gradient.  The adiabat given by \citet{Parmentier2015NongreyAnalyticalModel} which is a fit to \citet{Saumon1995EquationStateLowMass} is
\begin{equation}
    \nabla_\text{ad} = \left(\frac{\partial \ln T}{\partial \ln P}\right)_S = a - bT
\end{equation}
with $a=0.32$ and $b=1/30000$~K$^{-1}$. Assuming we can treat the partial derivatives as total and integrating by partial fractions we arrive at the following equation extrapolating the temperature to any pressure
\begin{equation}
    T(P) = \frac{a K(P)}{1 + bK(P)}
\end{equation}
where
\begin{equation}
    K = \frac{T_0}{a-bT_0}\left(\frac{P}{P_0}\right)^a
\end{equation}
and where $(T_0, P_0)$ is a known T-P point at the bottom of the un-extended atmosphere.  The eddy diffusion coefficient can then be extended as $K_{zz} \propto H^{-1/3}$ where $H = {kT}/{\mu m_p g}$ is the scale height \citep[][Eqn. 16]{Giersasch1985EnergyConversionProcesses}. 

\subsection{Post-processing}\label{sec:post}
\texttt{CARMApy} includes built in methods for reading in results and then generating spectra from the simulation output. \texttt{CARMApy} uses \texttt{pyFastchem} to calculate equilibrium gas phase abundances to serve as the gas phase input into \texttt{PICASO} \citep{Batalha2019ExoplanetReflectedlightSpectroscopy}\footnote{see \href{https://carmapy.readthedocs.io/en/latest/_autosummary/carmapy.Results.html\#carmapy.Results.gen_picaso_atm_file}{\texttt{carmapy.Results.gen\_picaso\_atm\_file()}}}.  
For the cloud phase the extinction coefficient, $\beta_\text{ext}$, scattering coefficient, $\beta_\text{sca}$, and the asymmetry parameter, $g$, for each layer at each wavelength is calculated as follows \citep[Eqns. 12.26 - 12.29,][]{Petty2006FirstCourseAtmospheric}
\begin{align}
    \beta_\text{ext} (z, \lambda) &= \sum_\text{all cloud species} \int_0^\infty \pi r^2 n(z) Q_\text{ext}(r,\lambda) dr\\
    \beta_\text{sca}(z, \lambda) &= \sum_\text{all cloud species} \int_0^\infty \pi r^2 n(z) Q_\text{sca}(r, \lambda)dr \\
    g(z, \lambda) &= \frac{1}{\beta_\text{sca}} \sum_\text{all cloud species} \int_0^\infty \pi r^2 n(z) Q_\text{sca}(r, \lambda) g(r, \lambda)dr
\end{align}
where $Q_\text{ext}$ and $Q_\text{sca}$ are the extinction and scattering efficiencies which, along with $g(r, \lambda)$, are calculated using \texttt{PyMieScatt} \citep{Sumlin2018RetrievingAerosolComplex}.  From these values we can easily calculate the single scattering albedo, $\omega = \beta_\text{sca}/\beta_\text{ext}$, and the optical depth of each layer, $\Delta \tau = \beta_\text{ext}\Delta{z}$, and generate the required cloud phase inputs into \texttt{PICASO}\footnote{see \href{https://carmapy.readthedocs.io/en/latest/_autosummary/carmapy.Results.html\#carmapy.Results.gen_picaso_cloud_file}{carmapy.Results.gen\_picaso\_cloud\_file()}}.  While we do not package \texttt{PICASO} directly into \texttt{CARMApy} as we want to avoid unneeded dependencies, we include in our tutorial sample scripts for generating spectra with \texttt{CARMApy} and \texttt{PICASO}. \added{Note that these calculations are done assuming homogeneous spheres neglecting the core material as small scale tests with a coated spheres treatment lead to limited differences in the produced spectra at the cost of substantial computational complexity.} 

\subsection{Example Outputs}\label{sec:example}

\begin{figure}
    \centering
    \includegraphics[width=0.9\linewidth]{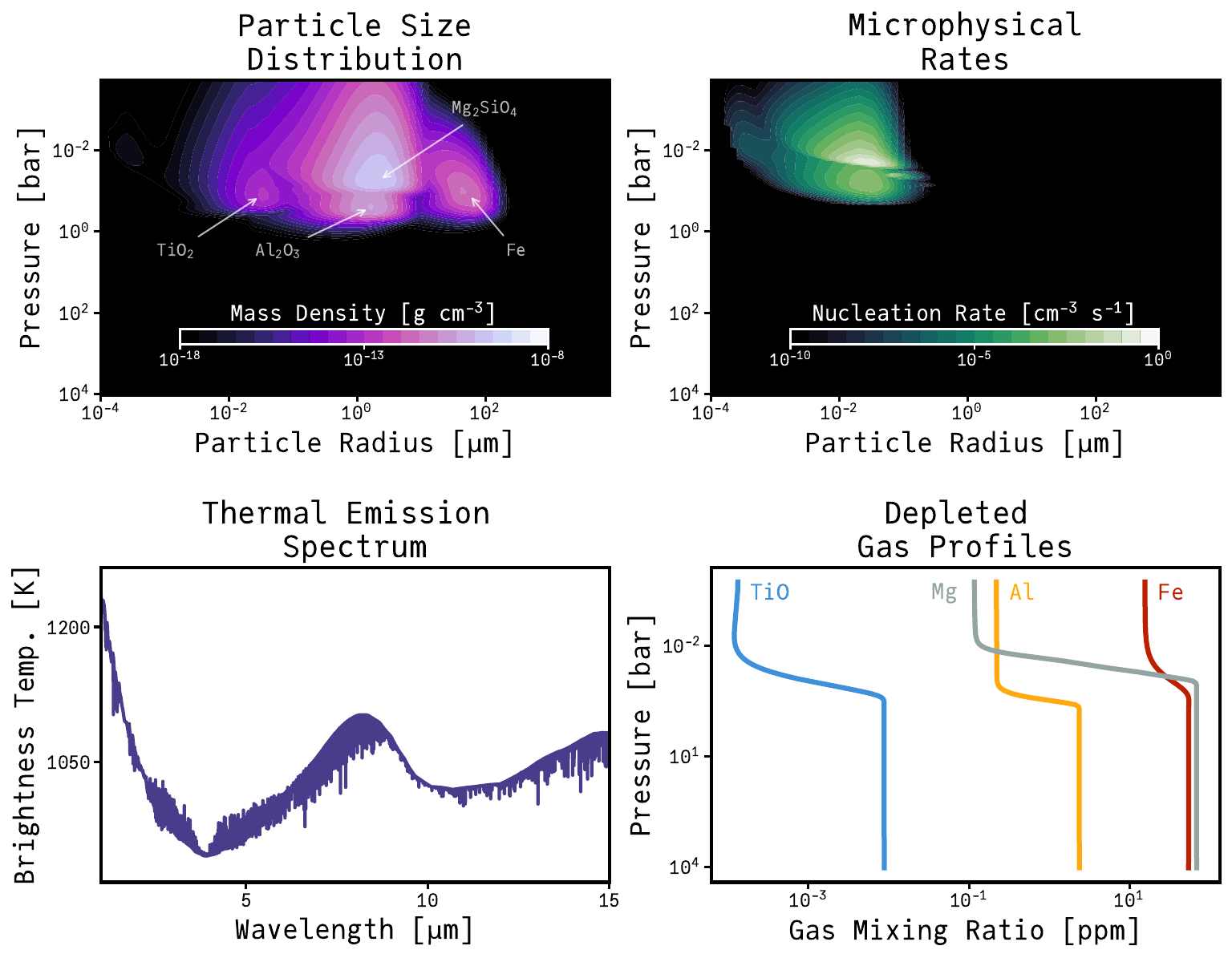}
    \caption{A selection of the data products that can be generated using \texttt{CARMApy}.   Example code to generate all of these type of plots are available in the \texttt{CARMApy} tutorials at \href{https://carmapy.readthedocs.io}{carmapy.readthedocs.io} \textbf{Top Left:} A cloud particle size distribution across the pressure and particle size grid generated by a 1D \texttt{CARMApy} run. \textbf{Top Right:} The heterogeneous nucleation rate for the same 1D run.  \texttt{CARMApy} is also able to output the homogeneous nucleation rate, the condensational growth rate, and evaporation rate. \textbf{Bottom Left:} A thermal emission spectrum generated by a 1D \texttt{CARMApy} run post processed with \texttt{PICASO}.  \textbf{Bottom Right:} The depleted gas abundance profiles generated by a \texttt{CARMApy} run.}
    \label{fig:example}
\end{figure}

To demonstrate the use of \texttt{CARMApy}, we present in Figure \ref{fig:example} a few example data products producible using the code.  Some example outputs from \texttt{CARMApy} include a cloud particle size distribution profile derived \textit{a priori} from our microphysical equations (Figure \ref{fig:example}, Top Left), the gain and loss rates in each grid cell due to the various microphysical processes (Figure \ref{fig:example}, Top Right), spectra generated using \texttt{PyMieScatt} \citep{Sumlin2018RetrievingAerosolComplex} and \texttt{PICASO} \citep{Batalha2019ExoplanetReflectedlightSpectroscopy}  (Figure \ref{fig:example}, Bottom Left), and the depleted gas abundance profiles caused by cloud formation (Figure \ref{fig:example}, Bottom Right).  All the 1D examples presented here are based on a 1800 K Brown Dwarf pressure, temperature, and eddy-diffusion structure taken from \citet{Morley2024SonoraSubstellarAtmosphere}.  \added{These example results show a large population of Mg$_2$SiO$_4$ centered on $\sim$1 $\mu$m, as well as other populations of TiO$_2$, Al$_2$O$_3$ and Fe.  While these runs also produce large silicate particles, we also produce a large number of sub-micron particles as observed by \citep{Burningham2021CloudBustingEnstatite}.  These silicates heterogeneously nucleate on TiO$_2$ at  $\sim$10 mbar and then grow to micron size while settling to $\sim$100 mbar.  In the thermal emission spectrum we can see a clear 10 micron silicate feature, although further analysis needs to be done to determine how well the shape of the feature matches observations.  Such analysis, along with a detailed study of cloud formation in brown dwarfs, is being undertaken in Cukier et al. (\textit{in prep.}). Tutorials on how to setup and run \texttt{CARMApy} and then generate these plots are available at \href{https://carmapy.readthedocs.io/en/v1.0.1/notebooks/1_my_first_carma.html}{https://carmapy.readthedocs.io/en/v1.0.1/notebooks/1
\_my\_first\_carma.html} and \href{https://carmapy.readthedocs.io/en/v1.0.1/notebooks/2_results.html}{https://carmapy.readthedocs.io/en/v1.0.1/notebooks/2\_results.html} respectively.}

\subsection{2D \texttt{CARMApy}}
\added{Planets are not one dimensional objects and thus multidimensional effects are often important when considering the cloud properties of one.  As such multi-dimensional models of planetary atmospheres are useful to understand these bodies, and such models, to be complete, ought to have a prescription for clouds.  Exoplanet GCMs have included clouds in a number of different prescriptions, such as using prescribed source terms and particle sizes to model the cold-trap on hot Jupiters \citep{Parmentier20133DMixingHot}, fixing cloud properties based on observations to understand how aerosols affect the heating and thus thermal structure of hot Jupiters \citep{Roman2017ModelingEffectsInhomogeneous}, and coupling GCMs directly with ECC formation models \citep{Christie2021ImpactMixingTreatments} or moment method kinetic cloud formation models \citep{Lee2023ModellingDynamicallyDriven}. These models all trade off resolving differing amounts of physics in exchange for varying levels of computational efficiency.}

\added{2D \texttt{CARMApy} maintains a complete bin-scheme microphysical treatment of cloud formation in exchange for having a less complete picture of transport around the planet.} In this mode, the longitudinally varying cloud profile is calculated at the equator by advecting the cloud column uniformly around the equator, holding the pressure coordinate of each bin constant while allowing the temperature at each pressure level to vary longitudinally. This process is done in log-pressure coordinates and introduces a vertical metric factor of $ds/dz = T/T_0$ where $T_0$ is the temperature at the base of the atmosphere \citep{Toon1988MultidimensionalModelAerosols}. The speed of advection is user specified, but typically has been chosen to be the average wind speed at the base of the cloud corresponding to the most relevant homogeneously nucleating condensate.  \added{This pseudo 2D method is similar to methods used in chemical models of exoplanet atmospheres \citep[eg.][]{Agundez2014Pseudo2DChemical, Baeyens2021GridPseudo2DChemistry}} and was first implemented in \texttt{ExoCARMA} by \citet{Powell2024TwodimensionalModelsMicrophysical}.  These 2D models are particularly suited to modeling planets which have strong longitudinal asymmetry and stable equatorial jets which dominate the observable region of the planetary atmosphere, such as hot Jupiters  \citep[eg. ][]{Showman2009AtmosphericCirculationExoplanets},\added{ and additionally can help resolve questions such as the non-local depletion of TiO and VO \citep{Roth2024HotJupiterDiversity}. }

\begin{figure}
    \centering
    \includegraphics[width=0.9\linewidth]{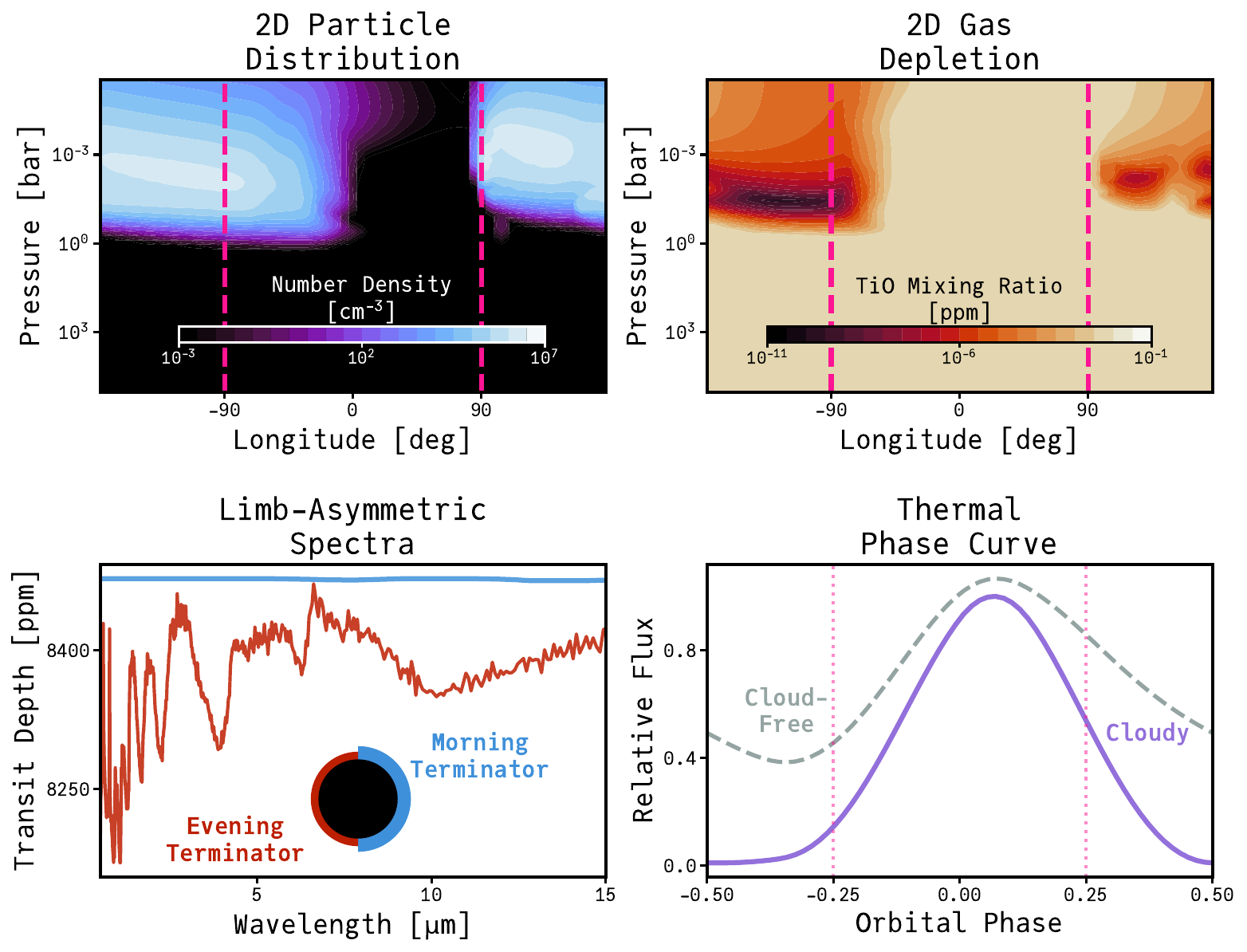}
    \caption{A selection of the data products that can be generated using 2D-\texttt{CARMApy}.  Example code to generate all of these type of plots are available in the \texttt{CARMApy} tutorials at \href{https://carmapy.readthedocs.io}{carmapy.readthedocs.io} \textbf{Top Left:} The 2D cloud particle distribution profile generated by a 2D \texttt{CARMApy} run.  The plot is centered on the substellar point and the pink dashed lines are the morning and evening terminators.  \textbf{Top Right:} The TiO$_2$ gas abundance profile from a 2D \texttt{CARMApy} run showing the disequilibrium gas depletion caused by cloud formation.   The plot is centered on the substellar point and the pink dashed lines are the morning and evening terminators. \textbf{Bottom Left:} NIR transmission spectra calculated at the morning (blue) and evening (red) terminators as derived from a 2D \texttt{CARMApy} run.  \textbf{Bottom Right:} NIR thermal emission phase curve as derived from as 2D \texttt{CARMApy} run, as well as the same run with the clouds ignored. }
    \label{fig:example_2d}
\end{figure}

Recent observations of the limb asymmetries between the morning and evening terminators of exoplanets show that differing cloud properties play a critical role in explaining the differences in spectra measured at each of these limbs \citep{Mukherjee2026CloudyMorningsClear}.  2D \texttt{CARMApy}, and the underlying \texttt{CARMA} model, in modeling the cloud column at each longitude, are able to resolve differences in clouds and thus predict limb-asymmetric spectra (see Samra et al. in prep, Kennedy et al. in prep).  Additionally this longitude resolution allows us to extract observables such as thermal emission phase curves from our models.  Example 2D \texttt{CARMApy} outputs and derived observables, including these limb asymmetric spectra and phase curves, are presented in Figure \ref{fig:example_2d}.  These outputs were generated using GCM data of an 1800 K Hot Jupiter from \citet{Roth2024LargeGridNonGrey}.  \added{From these models we can qualitatively reproduce observations of hot Jupiters such as the cloudy morning spectra and clear evening spectra seen in observations \citep[eg.][]{Mukherjee2026CloudyMorningsClear}, likely caused in part by the forsterite formed on the night side of the planet persisting into the day side, past where they would evaporate under equilibrium assumptions, and a similar yet reversed effect occurring at the evening terminator, as these microphysical processes take time. We also can qualitatively reproduce the larger phase curve amplitudes that are seen in cloudy planets \citep{Bell2024NightsideCloudsDisequilibrium}.  In this instance of the model, we do not see a cold trap of the TiO gas on the night side as predicted by \citet{Roth2024HotJupiterDiversity}, but we note that other instances of 2D \texttt{CARMApy} models not shown here do sometimes see this effect. A more full analysis of 2D \texttt{CARMApy} models of hot Jupiters will be published in D. Samra et al. (\textit{in prep}).}  Tutorials showing how these data products are created are available at \href{https://carmapy.readthedocs.io/en/v1.0.1/notebooks/5_2d_carmapy.html}{carmapy.readthedocs.io/en/latest/notebooks/5\_2d\_carmapy.html}.

\section{Benchmarks} \label{sec:benchmark}
To ensure correctness, we benchmark the results of \texttt{CARMApy} against that of \texttt{ExoCARMA 1.0}.  For the benchmark, we use a 1-D P-T and eddy-diffusion structure of a 2000 K brown dwarf with log g $=4.5$ and $f_\text{sed} = 4$ from \citet{Morley2024SonoraSubstellarAtmosphere}.  \added{This is a very similar brown dwarf to the one presented in Section \ref{sec:example} and we point the reader there for a further discussion on the potential observables from this model.}  The atmospheric viscosity, thermal conductivity, and specific heat were set to that of a pure H$_2$ atmosphere\footnote{note that the viscosity and thermal conductivity used was not the up-to-date values used as defaults but was instead set to be consistent with \citep{Gao2018MicrophysicsKClZnS}}.  For this benchmark we initialize a simple nucleation network, allowing TiO$_2$ to homogeneously condense and allowing Mg$_2$SiO$_4$ to heterogeneously condense upon the TiO$_2$ CCN.  The model was run at a timestep of 100~s for 24000 timesteps with minimal I/O to isolate the timing to that of the  microphysics calculations.  This process took \texttt{CARMApy} $\sim$50 seconds on 4 cores or $\sim$100 seconds single-threaded on a M4 Macbook Pro.  The same process took \texttt{ExoCARMA 1.0} $\sim$190 seconds on the same test\footnote{Single threaded as OpenMP was not implemented in \texttt{ExoCARMA 1.0}}.  \added{Note that these runs were performed without coagulation and with our sedimentation velocity fixes included into \texttt{ExoCARMA 1.0} \footnote{Despite significantly changing the sedimentation velocity functional form from \texttt{ExoCARMA 1.0} to match the literature and the form used in Earth \texttt{CARMA}, we find that such corrections led to negligible changes to the output}.}

\begin{figure}
    \centering
    \includegraphics[width=0.95\linewidth]{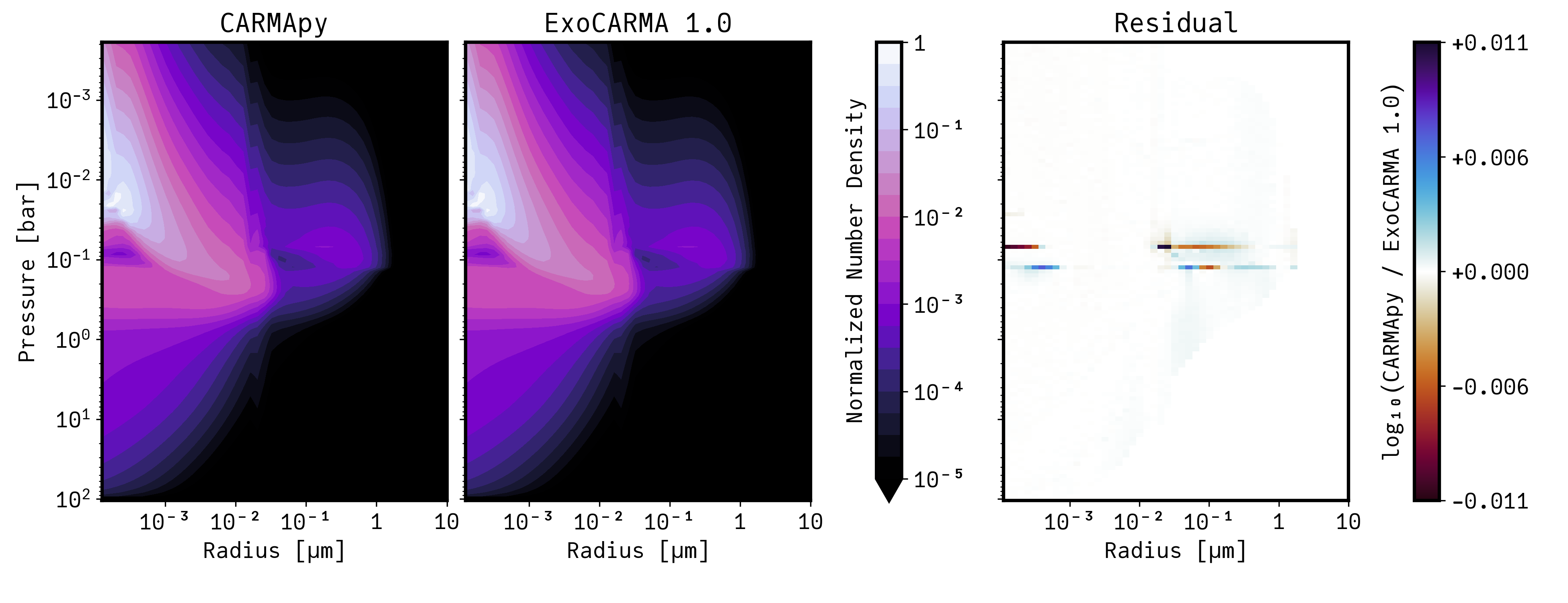}
    \caption{Comparison of the total condensate number density for a sample 2-species 2000 K brown dwarf run, averaged over 1000 samples.  The output from \texttt{CARMApy} on the left is benchmarked against the \texttt{ExoCARMA} code without our edits on the middle panel.  {The residual is presented on the panel on the right.  Note that the horizontal stripes are located at the cloud base and are substantially lower in magnitude than the temporal variability of the runs}}
    \label{fig:benchmark_sizdist}
\end{figure}

We then resumed those same simulations with outputting data every 10 timesteps for another 10000 timesteps.  We then averaged over these 1000 outputs as \texttt{CARMA} can fluctuate around equilibria but long term averages tend to be stable. The combined TiO$_2$ and Mg$_2$SiO$_4$ particle size distribution profiles, \added{and their associated residuals}, are presented Figure \ref{fig:benchmark_sizdist}.  These size distributions are, as expected, practically visually indistinguishable.  The total cloud and gas mass for each species agrees up to a relative difference $\sim$$3\times10^{-4}$ and $\sim$$3\times 10^{-5}$ respectively.  \added{On a bin-by-bin basis the typical amount a given altitude and radius bin varies between benchmarks is $\sim$$2\times 10^{-4}$ in mass density with the worst disagreements being about 1\%. Note that these regions of relatively high disagreement are localized to the cloud base and are still smaller than the temporal variability of the models.  Similarly the area weighted mean radius is consistent within $\sim$$1\times10^{-4}$ relative difference on average across all altitude levels and within $\sim$$1\times10^{-3}$ in the level with the worst agreement.}

\begin{figure}
    \centering
    \includegraphics[width=0.95\linewidth]{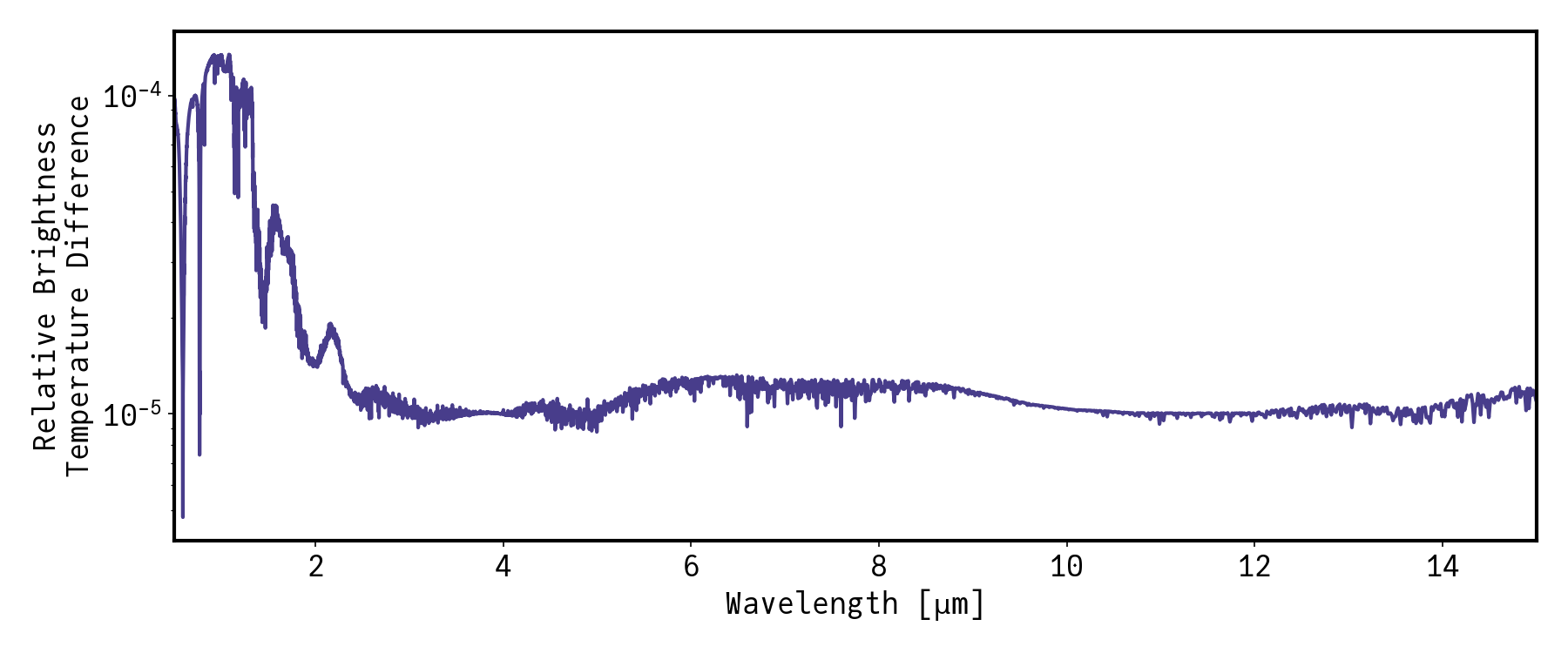}
    \caption{The relative difference in the brightness temperature emission spectra between \texttt{ExoCARMA 1.0} and \texttt{CARMApy} for a sample 2-species 2000 K brown dwarf run, with number densities averaged over 1000 samples.}
    \label{fig:benchmark_spectra}
\end{figure}

To get a sense of how these changes impact the observable features of these bodies, we present our analysis of our emission spectra benchmarking in Figure \ref{fig:benchmark_spectra}.  This figure shows the relative difference in the brightness temperature emission spectra between \texttt{ExoCARMA 1.0} and \texttt{CARMApy}.  The maximum absolute error is on the order of $0.3$~K which corresponds to a relative error of $\sim$$10^{-4}$.  At longer wavelengths the maximum absolute error drops to $\sim$$0.01$~K.  corresponding to a relative error of $\sim$$10^{-5}$.  These deviations are of a magnitude which is consistent with floating point error propagated over tens of thousands of timesteps and are small enough to be considered negligible as other assumptions, such as the choice of the eddy-diffusion structure of the atmosphere, will far dominate the uncertainty presented here.

\section{Future Development}

While \texttt{CARMApy} is already a powerful code, there remains improvements to be made for future releases.  Future releases of \texttt{CARMApy} likely will include some of the following features:
\begin{enumerate}
    \item A more complete solution for including hazes, including fractal aggregates and cloud-haze interactions \citep[see][]{Nagpal2026WetRemovalCloud}
    \item The ability to have more than one condensate species formed from a single limiting gas (ie. both SiO and SiO$_2$ forming from SiO gas)
    \item Radiatively active clouds
    \item Coupling with gas phase chemistry 
    \item Expanding the number of default atmospheric parameters and condensates available with \texttt{CARMApy}
\end{enumerate}
As we continue to use and refine \texttt{CARMApy} the solutions we develop will be released to the same channels as described herein.

\section{Conclusions}
Here we have presented the new python-based bin-scheme microphysical cloud model \texttt{CARMApy}, as well as an update to the underlying Fortran code, \texttt{ExoCARMA 2.0}.   \texttt{CARMApy} is \texttt{pip} installable, is available on \texttt{GitHub},\footnote{\href{https://github.com/wcukier/carmapy}{https://github.com/wcukier/carmapy}} and has tutorials and documentation hosted at \href{https://carmapy.readthedocs.io}{carmapy.readthedocs.io}.   The code and related documentation are also archived with Zenodo.\footnote{\href{https://doi.org/10.5281/zenodo.22693435}{https://doi.org/10.5281/zenodo.22693435} \citep{Cukier2026CARMApyOpenSourcePythona}} 

In addition to detailed microphysical modeling of exoplanets in both 1D and 2D,  \texttt{CARMApy} includes a number of features to help with the initialization and analysis of the model.  Interfaces with \texttt{pyFastchem} provide for straightforward setting of chemical equilibrium abundances.  Similarly, we provide post-processing code that exposes the simulation results in an intuitive interface, allows for quick plotting of size distributions, microphysical rates, and depleted gas profiles, and allows for fast and simple post-processing the output of the simulations into spectra with \texttt{PICASO}.  \texttt{CARMApy} also exposes interfaces such that the user is able to easily change the condensates included in the model as well as change the background atmospheric parameters. We have benchmarked our code against the old version of the Fortran code \texttt{ExoCARMA 1.0} and find that our updated code is able to reproduce the results of the old code to within some negligible deviations, while running $\sim$1.9 times faster single threaded and $\sim$3.8 times faster multithreaded.

\begin{acknowledgments}
The authors would like to thank Thomas Kennedy for helping beta test this code.  We would additionally like to thank the anonymous reviewer for their helpful comments. LLM tools (Sonnet 4.6, Opus 4.7-5) were used to format tables, to help implement and guide reasoning about code, format figures, and to suggest phrasing of sentences during preparation of this manuscript. This work benefited from the 2025 Exoplanet Summer Program in the Other Worlds Laboratory (OWL) at the University of California, Santa Cruz, a program funded by the Heising-Simons Foundation and NASA. X.Z. acknowledges support from the NSF grant (AST2307463), NASA Exoplanet Research grant (80NSSC22K0236), and the NASA Interdisciplinary Consortia for Astrobiology Research grant (80NSSC21K0597). VN acknowledges support from the National Science Foundation Graduate Research Fellowship under Grant No. DGE 2140001. D.S. acknowledges funding as part of JWST GO program 3969 (PIs: Espinoza, Powell).
\end{acknowledgments}

\begin{contribution}

WC led the writing of this paper, contributed to the development of \texttt{ExoCARMA}, and conceived of and led the development of the \texttt{CARMApy} wrapper.  DP co-led the development of \texttt{ExoCARMA} and supervised the writing of this paper and the development of the \texttt{CARMApy} wrapper.  XZ supervised and contributed to the development of \texttt{ExoCARMA}, supervised the development of the \texttt{CARMApy} wrapper, and helped edit this paper.  PG co-led the development of \texttt{ExoCARMA} and helped edit this paper.   DS and VN both contributed to the development of the \texttt{CARMApy} wrapper and helped edit this paper.


\end{contribution}

%

\software{\texttt{numpy} \citep{Harris2020ArrayProgrammingNumPy}, \texttt{matplotlib} \citep{Hunter2007Matplotlib2DGraphics}, \texttt{scipy} \citep{Virtanen2020SciPy10Fundamental}, \texttt{pyfastchem} \citep{Stock2022FASTCHEM2Improved}, \texttt{PyMieScatt} \citep{Sumlin2018RetrievingAerosolComplex},
\texttt{PICASO} \citep{Batalha2019ExoplanetReflectedlightSpectroscopy}, \texttt{cmasher} \citep{Velden2020CMasherScientificColormaps}, \citet{Petroff2024AccessibleColorSequences}, \texttt{CARMApy}  \citep{Cukier2026CARMApyUserFriendlyOpenSource}}



\bibliography{references}{}
\bibliographystyle{aasjournalv7}
\newpage
\appendix
\section{Table of Variables}
\startlongtable
\begin{deluxetable}{lp{11cm}}
\tabletypesize{\small}
\tablecaption{Summary of variables used in this work.\label{tab:variables}}
\tablewidth{0pt}
\tablehead{\colhead{Symbol} & \colhead{Description [units]}}
\startdata
$a_c$                               & Critical radius for nucleation [cm]\\
$A_k$                               & Kelvin factor [1]\\
$\alpha_0,\alpha_1,\alpha_2,\alpha_3$ & Coefficients of the saturation vapor pressure fit [1, K, 1, 1]\\
$\beta$                             & Interpolation factor between continuum and kinetic coagulation regimes [1] \\
$B_0,\ldots,B_6$                    & Empirical coefficients for the polynomial fit of $Re$ as a function of the Best number (Table~\ref{tab:fit_coeffs}) [1]\\
$c_p$                               & Specific heat capacity of the atmosphere [erg/g/K]\\
$c_\text{surf}$                     & Surface number density of condensate molecules on the CCN [cm$^{-2}$]\\
$C_c$                               & Cunningham slip factor [1]\\
$C_D$                               & Drag coefficient [1]\\
$d_q$                               & Collision diameter of the limiting gas molecule [cm]\\
$D_i$                               & Brownian diffusion coefficient of particle species $i$ [cm$^2$/s]\\
$D_v$                               & Molecular diffusion coefficient of the condensate vapor [cm$^2$/s]\\
$D_v'$                              & Corrected molecular diffusion coefficient [cm$^2$/s]\\
$\delta_i$                          & Transitional length scale for particle $i$ in the coagulation regime interpolation [cm]\\
$\Delta F_c$                        & Free energy of formation of a critical nucleus [erg]\\
$\Delta F''_c$                      & Second derivative of $\Delta F_c$ w.r.t.\ molecule number, evaluated at $a_c$ [erg]\\
$E_A$                               & Collision efficiency in the aerodynamic limit ($Re \gg 1$) [1]\\
$E_\text{coalesce}$                 & Coalescence (sticking) efficiency [1]\\
$E_\text{collide}$                  & Collision efficiency for gravitational coalescence [1]\\
$E_\text{Fuchs}$                    & Fuchs collision efficiency, correcting $E_\text{Langmuir}$ for direct interception by finite-size particles [1]\\
$E_\text{Langmuir}$                 & Langmuir collision efficiency [1]\\
$E_V$                               & Collision efficiency in the viscous limit ($Re \ll 1$) [1]\\
$\eta_a$                            & Dynamic viscosity of the atmosphere [cm$^2$/s]\\
$\eta_1,\eta_2,\eta_3,\eta_4$          & Coefficients for the dynamic viscosity parameterization [poise, 1, K, K$^2$]\\
$f$                                 & Shape factor relating heterogeneous to homogeneous free energy of formation [1]\\
$f_r$                               & Particle mass mixing ratio [1]\\
$f_\text{sed}$                      & Sedimentation efficiency parameter (equilibrium cloud condensation models) [1]\\
$[\text{Fe}/\text{H}]$              & Atmospheric metallicity [1]\\
$F_\text{ad}$                       & Advective flux across a bin boundary [cm$^{-2}$s$^{-1}$]\\
$F_\text{des}$                      & Desorption energy of the condensate molecule from the CCN [erg]\\
$F_\text{dif}$                      & Diffusive flux across a bin boundary [cm$^{-2}$s$^{-1}$]\\
$F_t$                               & Thermal ventilation factor [1]\\
$F_v$                               & Mass ventilation factor [1]\\
$g$                                 & Gravitational acceleration [cm/s$^2$]\\
$g_m$                               & Number of molecules in a particle of radius $a_c$ [1]\\
$H$                                 & Atmospheric pressure scale height, $H = kT/\mu m_p g$ [cm]\\
$J_\text{het}$                      & Heterogeneous nucleation rate [s$^{-1}$]\\
$J_\text{hom}$                      & Homogeneous nucleation rate [s$^{-1}$ cm$^{-3}$]\\
$J_{ki}$                            & Heterogeneous nucleation rate producing bin-$k$ particles on bin-$i$ CCN [s$^{-1}$]\\
$k$                                 & Boltzmann constant [erg/K]\\
$K^B_{12}$                          & Brownian coagulation kernel  [cm$^3$ s$^{-1}$]\\
$K^{DE}_{12}$                       & Convective diffusion enhancement coagulation kernel [cm$^3$ s$^{-1}$]\\
$K^G_{12}$                          & Gravitational collection coagulation kernel [cm$^3$ s$^{-1}$] \\
$K^\text{tot}_{12}$                 & Total coagulation kernel [cm$^3$ s$^{-1}$]\\
$K_{zz}$                            & Eddy (turbulent) diffusion coefficient [cm$^2$ s$^{-1}$]\\
$\kappa_a$                          & Thermal conductivity of the atmosphere [erg/cm/s/K]\\
$\kappa_a'$                         & Corrected atmospheric thermal conductivity [erg/cm/s/K]\\
$\kappa_0,\kappa_1,\kappa_2$        & Coefficients for the thermal conductivity parameterization [erg/s/cm/K, erg/s/cm/K$^2$, erg/s/cm/K$^3$]\\
$\text{Kn}$                         & Knudsen number for transport [1]\\
$\text{Kn}_c$                       & Collisional Knudsen number of the condensing gas [1]\\
$\text{Kn}_t$                       & Thermal Knudsen number of the condensing gas [1]\\
$L$                                 & Latent heat of vaporization [erg/g]; also the loss rate in the semi-implicit coagulation/growth update [s$^{-1}$]\\
$\ell,\,\ell_c,\,\ell_t$            & Mean free paths associated with $\text{Kn}$, $\text{Kn}_c$, and $\text{Kn}_t$, respectively [cm]\\
$\ell_i$                            & Mean free path of particle $i$ [cm]\\
$\lambda_c,\,\lambda_t$             & Correction coefficients for $D_v'$ and $\kappa_a'$ [1]\\
$m$                                 & Mass of a cloud particle [g]\\
$m_i$                               & Mass of particles in bin $i$ [g]\\
$m_{i+j}$                           & Combined mass of two coagulating particles, $m_{i+j} = m_i + m_j$ [g]\\
$m_p$                               & Proton mass [g]\\
$m_v$                               & Mass of a condensate gas particle [g]\\
$M$                                 & Molar mass of the condensate [g/mol]\\
$M_a$                               & Mean molar mass of the background atmosphere [g/mol]\\
$M_v$                               & Molar mass of the limiting (key) gas species [g/mol]\\
$\mu$                               & Cosine of the contact angle between condensate and CCN surface [1]\\
$n$                                 & Number density of condensate vapor molecules [cm$^{-3}$]\\
$n_k$                               & Number density of particles in size bin $k$ [cm$^{-3}$]\\
$\mathbf{n}^t$                      & State vector of all gas and particle bin number densities at timestep $t$ [cm$^{-3}$]\\
$N_A$                               & Avogadro's number [mol$^{-1}$]\\
$N_s$                               & Surface density of adsorption sites on the CCN [cm$^{-2}$]\\
$\nabla_\text{ad}$                  & Adiabatic temperature gradient, $(\partial \ln T / \partial \ln P)_S$ [1]\\
$\nu$                               & Attempt frequency for molecular desorption from the CCN [s$^{-1}$]\\
$\nu_\text{cloud}^\text{gas}$       & Stoichiometric ratio of the gas in the condensation reaction [1]\\
$\nu_r^\text{key}$                  & Stoichiometric ratio of the limiting reactant in a Type~III reaction [1]\\
$P$                                 & Atmospheric pressure [dyn/cm$^2$]\\
$P^\downarrow_{ijk},\,P^\uparrow_{ijk}$ & Partition factors distributing a coagulation product between adjacent mass bins [1]\\
$p_i$                               & Equilibrium partial pressure of gas species $i$ [dyn/cm$^2$]\\
$p'_i$                              & Saturation vapor pressure of gas species $i$ [dyn/cm$^2$]\\
$p_s$                               & Saturation vapor pressure of condensate [dyn/cm$^2$]\\
$p_v$                               & Partial pressure of condensate vapor [dyn/cm$^2$]\\
$\Phi$                              & Diffusion flux of vapor molecules [cm$^{-2}$s$^{-1}$]\\
$r$                                 & Radius of a cloud particle [cm]\\
$r_m$                               & Mass ratio between adjacent mass bins, $r_m = m_{k+1}/m_k$ [1]\\
$r_N$                               & Radius of the cloud condensation nucleus (CCN) [cm]\\
$R$                                 & Ideal gas constant [erg/mol/K]\\
$R_d$                               & Radial distance from the center of a cloud particle [cm]\\
$Re$                                & Reynolds number of a falling particle [1]\\
$\rho_a$                            & Atmospheric mass density [g/cm$^3$]\\
$\rho_p$                            & Bulk density of condensed particles [g/cm$^3$]\\
$\rho_s$                            & Saturation vapor density at the particle surface [g/cm$^3$]\\
$\rho_v$                            & Condensate vapor density [g/cm$^3$]\\
$S$                                 & Saturation ratio, $S = p_v / p_s$ [1]; also the source (production) term in the semi-implicit coagulation/growth update [cm$^{-3}$s$^{-1}$]\\
$S_\text{het evap}$                 & Source term from cores left behind by evaporating heterogeneous particles [cm$^{-3}$s$^{-1}$]\\
$S_k$                               & Stokes number [1]; also the production rate into bin $k$ in the semi-implicit update [cm$^{-3}$s$^{-1}$]\\
$S_r$                               & Effective supersaturation ratio for Type~III reactions, $S_r = S^{1/\nu_r^\text{key}}$ [1]\\
$\sigma$                            & Surface tension of the condensate--vapor interface [erg/cm$^2$]\\
$\sigma_0,\,\sigma_1$               & Coefficients of the linear surface-tension fit, $\sigma = \sigma_0 - \sigma_1 T$ [erg/cm$^2$, erg/cm$^2$/K]\\
$T$                                 & Atmospheric temperature [K]\\
$T_p$                               & Temperature of the cloud particle [K]\\
$u$                                 & Advection velocity in the generalized coordinate $\xi$ [$\xi$/s]\\
$v_{ed}$                            & Eddy-diffusion vertical velocity [cm/s]\\
$v_i$                               & Mean thermal speed of particles in bin $i$ [cm/s]\\
$v_t$                               & Terminal (sedimentation) velocity of a particle [cm/s]\\
$v_\text{winds}$                    & User-specified upward wind speed [cm/s]\\
$v^\uparrow,\,v^\downarrow$         & Upward / downward effective velocities across a bin boundary [cm/s]\\
$\xi$                               & Generalized advection coordinate (altitude or particle mass) [cm or g]\\
$Z$                                 & Zeldovich factor (non-equilibrium correction to nucleation rate) [1]\\
$Z_\text{het}$                      & Zeldovich factor for heterogeneous nucleation [1]\\
\enddata
\end{deluxetable}

\section{Default \texttt{CARMApy} Condensates}\label{sec:defaults}
We present here the parameters used for the default \texttt{CARMApy} condensates in Tables \ref{tab:first} - \ref{tab:last}.  The indices of refraction for these condensates were taken from the sources in the \texttt{POSEIDON} \citep{MacDonald2017HD209458bNew, MacDonald2023POSEIDONMultidimensionalAtmospheric} opacity database \citep{Mullens2024ImplementationAerosolMie}.  The individual sources used for each condensate are cited in the respective tables.


\begin{deluxetable}{l l c l}\label{tab:first}
\tablecaption{Physical and Thermodynamic Parameters for KCl Condensate}
\tablehead{\colhead{Symbol} & \colhead{Description} & \colhead{Value} & \colhead{Reference}}
\startdata
$\rho_p$ & Condensed Density (g\,cm$^{-3}$) & 1.988 & -- \\
$M$ & Molecular Weight & 74.5 & -- \\
$\sigma_0$ & Surface Energy (erg\,cm$^{-2}$) & 179.52 & \citet{Kaye2005225SurfaceTensions} \\
$\sigma_1$ & Surface Energy Slope (erg\,cm$^{-2}$\,K$^{-1}$) & 0.07 &  \\
$\alpha_0$ & Saturation Vapor Pressure Offset & 7.6106 & \citet{Morley2012NeglectedCloudsDwarf} \\
$\alpha_1$ & SVP Temperature Coefficient (K)& 11382 &  \\
$\alpha_2$ & SVP Metallicity Coefficient & 0 &  \\
$\alpha_3$ & SVP Pressure Coefficient & 0 &  \\
 & Condensation Reaction & KCl = KCl [s] & \citet{Morley2012NeglectedCloudsDwarf} \\
 & Key (Limiting) Species & KCl & -- \\
$d_q$ & Collision Diameter (cm) & $3.08\times10^{-8}$ & \shortstack{Estimated from \\ \citet{Sanderson1976ChemicalBondsBond}} \\
$M_v$ & Molecular Weight of Limiting Species & 74.5 & -- \\
 & Indices of Refraction & -- & \shortstack{\citet{Wakeford2015TransmissionSpectralProperties} \\ \citet{Palik1985HandbookOpticalConstants}} \\
\enddata 
\end{deluxetable}

\begin{deluxetable}{l l c l}
\tablecaption{Physical and Thermodynamic Parameters for ZnS Condensate}
\tablehead{\colhead{Symbol} & \colhead{Description} & \colhead{Value} & \colhead{Reference}}
\startdata
$\rho_p$ & Condensed Density (g\,cm$^{-3}$) & 4.04 & -- \\
$M$ & Molecular Weight & 97.474 & -- \\
$\sigma_0$ & Surface Energy (erg\,cm$^{-2}$) & 860 & \citet{Zhang2003MolecularDynamicsSimulations} \\
$\sigma_1$ & Surface Energy Slope (erg\,cm$^{-2}$\,K$^{-1}$) & 0 &  \\
$\alpha_0$ & Saturation Vapor Pressure Offset & 12.812 & \citet{Morley2012NeglectedCloudsDwarf} \\
$\alpha_1$ & SVP Temperature Coefficient (K)& 15873 &  \\
$\alpha_2$ & SVP Metallicity Coefficient & 1 &  \\
$\alpha_3$ & SVP Pressure Coefficient & 0 &  \\
 & Condensation Reaction & H$_2$S + Zn $=$ ZnS[s] &  \citet{Morley2012NeglectedCloudsDwarf}\\
 & Key (Limiting) Species & Zn & -- \\
$d_q$ & Collision Diameter (cm) & $3.67\times10^{-8}$ & \shortstack{Estimated from \\ \citet{Zack2009PureRotationalSpectrum}} \\
$M_v$ & Molecular Weight of Limiting Species & 65.38 & -- \\
 & Indices of Refraction & -- & \shortstack{\citet{Wakeford2015TransmissionSpectralProperties} \\ \citet{Querry1987OpticalConstantsMinerals}} \\
\enddata
\end{deluxetable}

\begin{deluxetable}{l l c l}
\tablecaption{Physical and Thermodynamic Parameters for Na$_2$S Condensate}
\tablehead{\colhead{Symbol} & \colhead{Description} & \colhead{Value} & \colhead{Reference}}
\startdata
$\rho_p$ & Condensed Density (g\,cm$^{-3}$) & 1.856 & -- \\
$M$ & Molecular Weight & 78.0452 & -- \\
$\sigma_0$ & Surface Energy (erg\,cm$^{-2}$) & 1033 & E.\ Lee (priv.\ comm.) \\
$\sigma_1$ & Surface Energy Slope (erg\,cm$^{-2}$\,K$^{-1}$) & 0 &  \\
$\alpha_0$ & Saturation Vapor Pressure Offset & 8.55 & \citet{Morley2012NeglectedCloudsDwarf}\\
$\alpha_1$ & SVP Temperature Coefficient (K)& 13889 &  \\
$\alpha_2$ & SVP Metallicity Coefficient & 0.5 &  \\
$\alpha_3$ & SVP Pressure Coefficient & 0 &  \\
 & Condensation Reaction & H$_2$S + 2Na = Na$_2$S[s] + H$_2$ & \citet{Morley2012NeglectedCloudsDwarf} \\
 & Key (Limiting) Species & Na & -- \\
$d_q$ & Collision Diameter (cm) & $4.2\times10^{-8}$ & \shortstack{Estimated from\\ \citet{Glasser2000LatticeEnergiesUnit}} \\
$M_v$ & Molecular Weight of Limiting Species & 22.9898 & -- \\
 & Indices of Refraction & -- & \shortstack{\citet{Wakeford2015TransmissionSpectralProperties}, \\ \citet{Morley2012NeglectedCloudsDwarf}, \\ \citet{Montaner1979OpticalConstantsSodium}, \\\citet{Khachai2009FPAPW+loCalculationsElectronic}} \\
\enddata
\end{deluxetable}

\begin{deluxetable}{l l c l}
\tablecaption{Physical and Thermodynamic Parameters for MnS Condensate}
\tablehead{\colhead{Symbol} & \colhead{Description} & \colhead{Value} & \colhead{Reference}}
\startdata
$\rho_p$ & Condensed Density (g\,cm$^{-3}$) & 4.0 & -- \\
$M$ & Molecular Weight & 87.003 & -- \\
$\sigma_0$ & Surface Energy (erg\,cm$^{-2}$) & 2326 & E. Lee (priv. comm.) \\
$\sigma_1$ & Surface Energy Slope (erg\,cm$^{-2}$\,K$^{-1}$) & 0 &  \\
$\alpha_0$ & Saturation Vapor Pressure Offset & 11.532 & \citet{Morley2012NeglectedCloudsDwarf} \\
$\alpha_1$ & SVP Temperature Coefficient (K)& 23810 &  \\
$\alpha_2$ & SVP Metallicity Coefficient & 1 &  \\
$\alpha_3$ & SVP Pressure Coefficient & 0 &  \\
 & Condensation Reaction & H$_2$S + 2Mn = Mn$_2$S[s] + H$_2$ & \citet{Morley2012NeglectedCloudsDwarf} \\
 & Key (Limiting) Species & Mn & -- \\
$d_q$ & Collision Diameter (cm) & $3.68\times10^{-8}$ & \shortstack{Estimated from \\\citet{Honsberg2007MnS}} \\
$M_v$ & Molecular Weight of Limiting Species & 54.938 & -- \\
 & Indices of Refraction & -- & \shortstack{\citet{Kitzmann2018OpticalPropertiesPotential}, \\ \citet{Montaner1979OpticalConstantsSodium}, \\ \citet{Huffman1967OpticalPropertiesAMnS}} \\
\enddata
\end{deluxetable}

\begin{deluxetable}{l l c l}
\tablecaption{Physical and Thermodynamic Parameters for Cr Condensate}
\tablehead{\colhead{Symbol} & \colhead{Description} & \colhead{Value} & \colhead{Reference}}
\startdata
$\rho_p$ & Condensed Density (g\,cm$^{-3}$) & 7.15 & -- \\
$M$ & Molecular Weight & 51.9961 & -- \\
$\sigma_0$ & Surface Energy (erg\,cm$^{-2}$) & 2068.63 & \citet{Kaye2005225SurfaceTensions} \\
$\sigma_1$ & Surface Energy Slope (erg\,cm$^{-2}$\,K$^{-1}$) & 0.2 &  \\
$\alpha_0$ & Saturation Vapor Pressure Offset & 7.49 & \citet{Morley2012NeglectedCloudsDwarf} \\
$\alpha_1$ & SVP Temperature Coefficient (K)& 20592 &  \\
$\alpha_2$ & SVP Metallicity Coefficient & 0 &  \\
$\alpha_3$ & SVP Pressure Coefficient & 0 &  \\
 & Condensation Reaction & Cr  = Cr[s] & \citet{Morley2012NeglectedCloudsDwarf} \\
 & Key (Limiting) Species & Cr & -- \\
$d_q$ & Collision Diameter (cm) & $3.66\times10^{-8}$ & Atomic Radius \\
$M_v$ & Molecular Weight of Limiting Species & 51.9961 & -- \\
 & Indices of Refraction & -- & \shortstack{\citet{Kitzmann2018OpticalPropertiesPotential},\\ \citet{Lynch1997IntroductionDataSeveral}, \\ \citet{Rakic1998OpticalPropertiesMetallic}} \\
\enddata
\end{deluxetable}

\begin{deluxetable}{l l c l}
\tablecaption{Physical and Thermodynamic Parameters for Mg$_2$SiO$_4$ Condensate}
\tablehead{\colhead{Symbol} & \colhead{Description} & \colhead{Value} & \colhead{Reference}}
\startdata
$\rho_p$ & Condensed Density (g\,cm$^{-3}$) & 3.21 & -- \\
$M$ & Molecular Weight & 140.69 & -- \\
$\sigma_0$ & Surface Energy (erg\,cm$^{-2}$) & 436 & \citet{Kozasa1989FormationDustGrains} \\
$\sigma_1$ & Surface Energy Slope (erg\,cm$^{-2}$\,K$^{-1}$) & 0 &  \\
$\alpha_0$ & Saturation Vapor Pressure Offset & 14.88 & C. Visscher (priv. comm.) \\
$\alpha_1$ & SVP Temperature Coefficient (K)& 32488 &  \\
$\alpha_2$ & SVP Metallicity Coefficient & 1.4 &  \\
$\alpha_3$ & SVP Pressure Coefficient & 0.2 &  \\
 & Condensation Reaction & \shortstack{2Mg + SiO + 3H$_2$O\\ $\qquad$= Mg$_2$SiO$_4$[s] + 3H$_2$} & \citet{Helling2006DustBrownDwarfs} \\
 & Key (Limiting) Species & Mg & -- \\
$d_q$ & Collision Diameter (cm) & $3.85\times10^{-8}$ & \shortstack{Estimated from \\ \citet{Glasser2000LatticeEnergiesUnit}} \\
$M_v$ & Molecular Weight of Limiting Species & 24.305 & -- \\
 & Indices of Refraction & -- & \shortstack{\citet{Burningham2021CloudBustingEnstatite}, \\ \citet{Jager2003StepsInterstellarSilicate}} \\
\enddata
\end{deluxetable}

\begin{deluxetable}{l l c l}
\tablecaption{Physical and Thermodynamic Parameters for Fe Condensate}
\tablehead{\colhead{Symbol} & \colhead{Description} & \colhead{Value} & \colhead{Reference}}
\startdata
$\rho_p$ & Condensed Density (g\,cm$^{-3}$) & 7.87 & -- \\
$M$ & Molecular Weight & 55.845 & -- \\
$\sigma_0$ & Surface Energy (erg\,cm$^{-2}$) & 2565.2285 & \citet{Kaye2005225SurfaceTensions} \\
$\sigma_1$ & Surface Energy Slope (erg\,cm$^{-2}$\,K$^{-1}$) & 0.39 &  \\
$\alpha_0$ & Saturation Vapor Pressure Offset & 7.23 & \citet{Visscher2010ATMOSPHERICCHEMISTRYGIANT} \\
$\alpha_1$ & SVP Temperature Coefficient (K)& 20995 & \\
$\alpha_2$ & SVP Metallicity Coefficient & 0 &  \\
$\alpha_3$ & SVP Pressure Coefficient & 0 &  \\
 & Condensation Reaction & Fe = Fe[s] & \citet{Helling2006DustBrownDwarfs} \\
 & Key (Limiting) Species & Fe & -- \\
$d_q$ & Collision Diameter (cm) & $3.7\times10^{-8}$ & Atomic Radius \\
$M_v$ & Molecular Weight of Limiting Species & 55.845 & -- \\
 & Indices of Refraction & -- & \shortstack{\citet{Kitzmann2018OpticalPropertiesPotential},\\ \citet{Lynch1997IntroductionDataSeveral}} \\
\enddata
\end{deluxetable}

\begin{deluxetable}{l l c l}
\tablecaption{Physical and Thermodynamic Parameters for TiO$_2$ Condensate}
\tablehead{\colhead{Symbol} & \colhead{Description} & \colhead{Value} & \colhead{Reference}}
\startdata
$\rho_p$ & Condensed Density (g\,cm$^{-3}$) & 4.25 & -- \\
$M$ & Molecular Weight & 79.866 & -- \\
$\sigma_0$ & Surface Energy (erg\,cm$^{-2}$) & 535.124 & \citet{Lee2015DustBrownDwarfs} \\
$\sigma_1$ & Surface Energy Slope (erg\,cm$^{-2}$\,K$^{-1}$) & 0.04396 &  \\
$\alpha_0$ & Saturation Vapor Pressure Offset & 9.5489 & \citet{Helling2001DustBrownDwarfs} \\
$\alpha_1$ & SVP Temperature Coefficient (K)& 32456.8678 &  \\
$\alpha_2$ & SVP Metallicity Coefficient & 0 &  \\
$\alpha_3$ & SVP Pressure Coefficient & 0 &  \\
 & Condensation Reaction & TiO$_2$ = TiO$_2$[s] & \citet{Helling2006DustBrownDwarfs} \\
 & Key (Limiting) Species & TiO$_2$ & -- \\
$d_q$ & Collision Diameter (cm) & $3.39\times10^{-8}$ & \shortstack{Estimated from \\ \citet{Helling2001DustBrownDwarfs}} \\
$M_v$ & Molecular Weight of Limiting Species & 79.866 & -- \\
 & Indices of Refraction & -- & \shortstack{\citet{Lee20223DRadiativeTransfer}, \\ \citet{Ribarsky1997TitaniumDioxideTiO2}, \\ \citet{Zeidler2011NearinfraredAbsorptionProperties}} \\
\enddata
\end{deluxetable}

\begin{deluxetable}{l l c l}
\tablecaption{Physical and Thermodynamic Parameters for Al$_2$O$_3$ Condensate}
\tablehead{\colhead{Symbol} & \colhead{Description} & \colhead{Value} & \colhead{Reference}}
\startdata
$\rho_p$ & Condensed Density (g\,cm$^{-3}$) & 3.99 & -- \\
$M$ & Molecular Weight & 101.96 & -- \\
$\sigma_0$ & Surface Energy (erg\,cm$^{-2}$) & 690 & \citet{Kozasa1989FormationDustGrains} \\
$\sigma_1$ & Surface Energy Slope (erg\,cm$^{-2}$\,K$^{-1}$) & 0 &  \\
$\alpha_0$ & Saturation Vapor Pressure Offset & 17.7 & \citet{Wakeford2017HightemperatureCondensateClouds} \\
$\alpha_1$ & SVP Temperature Coefficient (K)& 45892.6 &  \\
$\alpha_2$ & SVP Metallicity Coefficient & 1.66 &  \\
$\alpha_3$ & SVP Pressure Coefficient & 0 &  \\
 & Condensation Reaction & 2Al + 3H$_2$O = Al$_2$O$_3$ + 3H$_2$ & \citet{Sedlmayr2014PhysicsChemistryCircumstellar}. \\
 & Key (Limiting) Species & Al & -- \\
$d_q$ & Collision Diameter (cm) & $3.83\times10^{-8}$ & \shortstack{Estimated From \\ \citet{Dobrovinskaya2009PropertiesSapphire}} \\
$M_v$ & Molecular Weight of Limiting Species & 26.98 & -- \\
 & Indices of Refraction & -- & \shortstack{\citet{Kitzmann2018OpticalPropertiesPotential}, \\ \citet{Begemann1997AluminumOxideOpacity}, \\ \citet{Koike1995ExtinctionSpectraCorundum}} \\
\enddata
\end{deluxetable}

\begin{deluxetable}{l l c l}\label{tab:last}
\tablecaption{Physical and Thermodynamic Parameters for SiO Condensate}
\tablehead{\colhead{Symbol} & \colhead{Description} & \colhead{Value} & \colhead{Reference}}
\startdata
$\rho_p$ & Condensed Density (g\,cm$^{-3}$) & 2.13 & -- \\
$M$ & Molecular Weight & 44.0849 & -- \\
$\sigma_0$ & Surface Energy (erg\,cm$^{-2}$) & 500 & \citet{Sedlmayr2014PhysicsChemistryCircumstellar} \\
$\sigma_1$ & Surface Energy Slope (erg\,cm$^{-2}$\,K$^{-1}$) & 0 &  \\
$\alpha_0$ & Saturation Vapor Pressure Offset & 14.12 & \citet{Gail2013SeedParticleFormation} \\
$\alpha_1$ & SVP Temperature Coefficient (K)& 21506.3 &  \\
$\alpha_2$ & SVP Metallically Coefficient & 0 &  \\
$\alpha_3$ & SVP Pressure Coefficient & 0 &  \\
 & Condensation Reaction & SiO = SiO[s] & \citet{Sedlmayr2014PhysicsChemistryCircumstellar} \\
 & Key (Limiting) Species & SiO & -- \\
$d_q$ & Collision Diameter (cm) & $4.04\times10^{-8}$ & \shortstack{Estimated from \\ \citet{Sedlmayr2014PhysicsChemistryCircumstellar}} \\
$M_v$ & Molecular Weight of Limiting Species & 44.0849 & -- \\
 & Indices of Refraction & -- & \shortstack{\citet{Kitzmann2018OpticalPropertiesPotential}, \\ \citet{Phillip1997SiliconMonoxideSiO}, \citet{Wetzel2013LaboratoryMeasurementOptical}} \\
\enddata
\end{deluxetable}

\added{
\begin{deluxetable}{l l c l}\label{tab:last}
\tablecaption{Physical and Thermodynamic Parameters for MgSiO$_3$ Condensate}
\tablehead{\colhead{Symbol} & \colhead{Description} & \colhead{Value} & \colhead{Reference}}
\startdata
$\rho_p$ & Condensed Density (g\,cm$^{-3}$) & 3.2 & -- \\
$M$ & Molecular Weight & 100.3887 & -- \\
$\sigma_0$ & Surface Energy (erg\,cm$^{-2}$) & 400 & \citet{Sedlmayr2014PhysicsChemistryCircumstellar} \\
$\sigma_1$ & Surface Energy Slope (erg\,cm$^{-2}$\,K$^{-1}$) & 0 &  \\
$\alpha_0$ & Saturation Vapor Pressure Offset & 13.43 & \citet{Batalha2025CondensationCloudsSubstellar} \\
$\alpha_1$ & SVP Temperature Coefficient (K)& 28665.0 &  \\
$\alpha_2$ & SVP Metallically Coefficient & 1 &  \\
$\alpha_3$ & SVP Pressure Coefficient & 0 &  \\
 & Condensation Reaction & Mg + SiO + 2H$_2$O = MgSiO$_3$ + 2 H$_2$ & \citet{Sedlmayr2014PhysicsChemistryCircumstellar} \\
 & Key (Limiting) Species & SiO & -- \\
$d_q$ & Collision Diameter (cm) & $4.65\times 10^{-8}$ & \shortstack{Estimated from \\ \citet{Sedlmayr2014PhysicsChemistryCircumstellar}} \\
$M_v$ & Molecular Weight of Limiting Species & 44.0849 & -- \\
 & Indices of Refraction & -- & \shortstack{\citet{Wakeford2015TransmissionSpectralProperties} \\ \citet{Egan1975InterstellarMediumUV} \\ \citet{Dorschner1995StepsInterstellarSilicate}} \\
\enddata
\end{deluxetable}}



\end{document}